\documentclass[review,table]{elsarticle}
\usepackage{amssymb}
\usepackage{amsmath}

\usepackage[T1]{fontenc}
\usepackage{amssymb}
\usepackage{wasysym}
\usepackage{setspace}
\usepackage{float}
\usepackage{caption}
\usepackage{booktabs}
\usepackage{changepage}
\usepackage[table]{xcolor}
\usepackage{multirow}
\usepackage{graphicx}  
\usepackage{url}
\usepackage{todonotes}
\usepackage{pifont}
\usepackage{xspace}
\usepackage{algorithm}
\usepackage{algpseudocode}
\usepackage{hyperref}

\usepackage{makecell}
\usepackage{todonotes}
\usepackage{amsmath,amsfonts}
\usepackage{booktabs}
\usepackage{tikz}
\usepackage{subcaption}
\usepackage{times}

\usepackage{siunitx}

\DeclareMathAlphabet{\pazocal}{OMS}{zplm}{m}{n}
\definecolor{lightyellow}{rgb}{1.0, 1.0, 0.88}
\definecolor{lightblue}{rgb}{0.68, 0.85, 0.90}

\newtheorem{definition}{Definition}[section]

\definecolor{lightyellow}{rgb}{1.0, 1.0, 0.88}
\definecolor{lightblue}{rgb}{0.68, 0.85, 0.90}

\newcommand{\ME}{\mathcal{E}}
\newcommand{\MA}{\mathcal{A}}

\newcommand{\MT}{\mathcal{T}}

\newcommand{\MD}{\mathcal{V}}
\newcommand{\MN}{\mathcal{N}}
\newcommand{\MP}{\mathcal{P}}
\newcommand{\ML}{\mathcal{L}}
\newcommand{\MV}{\mathcal{V}}

\newcommand{\MS}{\mathcal{S}}
\newcommand{\MR}{\mathcal{R}}
\newcommand{\unif}{\pazocal{U}}
\newcommand{\MI}{\mathcal{I}}

\journal{Information Systems}

\begin{document}

\begin{frontmatter}



\title{An Experimental Comparison of Alternative Techniques for Event-Log Augmentation}


\author{Alessandro Padella\fnref{label1}}\ead{alessandro.padella@unipd.it}
\author{Francesco Vinci\fnref{label1}}\ead{francesco.vinci@unipd.it} 
\author{Massimiliano de Leoni\fnref{label1}}\ead{deleoni@math.unipd.it} 

\affiliation[label1]{organization={University of Padua},
            addressline={Via Trieste 63}, 
            city={Padua},
            postcode={35121}, 
            country={Italy}}

\begin{abstract}
Process mining analyzes and improves processes by examining transactional data stored in event logs, which record sequences of events with timestamps. However, the effectiveness of process mining, especially when combined with machine or deep learning, depends on having large event logs. Event log augmentation addresses this limitation by generating additional traces that simulate realistic process executions while considering various perspectives like time, control-flow, workflow, resources, and domain-specific attributes.
Although prior research has explored event-log augmentation techniques, there has been no comprehensive comparison of their effectiveness.
This paper reports on an evaluation of seven state-of-the-art augmentation techniques across eight event logs. The results are also compared with those obtained by a baseline technique based on a stochastic transition system. The comparison has been carried on analyzing four different aspects: similarity, preservation of predictive information,  information loss/enhancement, and computational times required. Results show that, considering the different criteria, a technique based on a stochastic transition system combined with resource queue modeling would provide higher quality synthetic event logs.
Event-log augmentation techniques are also compared with traditional data-augmentation techniques, showing that the former provide significant benefits, whereas the latter fail to consider process constraints.
\end{abstract}




\end{frontmatter}



\section{Introduction}\label{sec:introduction}\noindent
Process mining seeks to analyze and improve processes by analyzing process' transactional data, which reports on how individual executions are performed~\cite{AalstBook}. Process transaction data are organized in so-called \textit{event logs}, containing a collection of traces, each referring to an individual process' execution and is constituted by a sequence of events, which in turn record the starting or completion of given process activities within the process execution along with the timestamps in which they occurred.

Process mining can only be applied if an event log is available. As reported by Zimmermann et al. in~\cite{zimmermann}, data availability remains one of the most critical obstacles for process mining experts, both for practitioners and academics. The lack of availability of a sufficient amount of data is particularly challenging for those techniques that are based on machine- and deep-learning models that are very ``data greedy”. Examples in process mining refer to simulation and predictive process monitoring and/or prescriptive analytics~\cite{Mehdiyev2021, DiFrancescomarino2022}. Typically, these ``data greedy” techniques leverage on artificial intelligence-based models (e.g., neural networks) that require a large amount of data to be properly trained. It follows that, if the original event log is limited in size, these techniques cannot be successfully employed. This motivates the needs for event-log augmentation that can extend the original collections of traces with new ones.

\noindent The problem of augmenting an existing event log is delineated as follows:
\begin{quote}
\textit{Given an event log $\ML$ consisting of a collection of traces related to executions of a process $\MP$, the event-log augmentation aims to generate a novel collection of $N$ traces that record the potential events of new valid executions of $\MP$, where value $N$ is set by the process analysts.}
\end{quote}

\noindent
Event-log augmentation should be aimed at generating events that cover multiple perspectives. Indeed, events occur at certain timestamps (\textit{time} perspective) and refer to the execution of activities within individual process' executions (\textit{control-flow} perspective) that are executed by resources that play certain roles (\textit{resource/organizational} perspective). In addition to these general features, shared among almost every event-log, each event-log often includes domain-specific attributes related to the context of the log, which should also be considered and generated in the augmentation process (\textit{attribute} perspective).

The problem of a lack of enough amount of data is not novel, and techniques have been proposed in the artificial intelligence (AI) literature to augment the original data (cf.\ Section~\ref{sec:related_works_augmentation}). However, traditional problems of data augmentation rely on the assumption that data samples are stochastically independent~\cite{data_survey}. This assumption does not hold when one aims to augment an event log. Since there are constraints on the order of activities, the next events in a trace may heavily depend on the events that precede, as well as the occurrence of certain events may also depend on the events of other traces. Furthermore, events refer to activities performed by resources, and resources are shared, consequently, events from different traces also influence each other.

Note that event-log augmentation can also specifically focus on generating additional traces that record rare process executions. This allows upsampling of the event logs to increase the frequency of otherwise rare behavior. For instance, this is very useful when, in predictive process monitoring scenarios, one wants to accurately predict whether or not certain rare behavior is going to be observed in the future for running process executions (e.g., whether or not a rare, undesired activity is observed).

This article contributes to the field of event-log augmentation by providing an extensive evaluation of seven techniques suitable for event-log augmentation. 

To ensure a comprehensive and balanced evaluation, we introduce a baseline method that relies exclusively on an annotated transition system, in contrast to the other approaches that are all grounded in deep learning. This baseline not only offers a fundamentally different, model-driven perspective on event-log augmentation but also serves as a reference point to better assess the strengths, limitations, and added value of learning-based techniques.

Eight different event logs have been used for comparison, which are based on metrics that analyze various aspects. i) \textbf{Similarity} between real and generated event-logs has been assessed considering five different aspects: control-flow, time, congestion, resource and attributes generation. This have been analyzed employing the metrics proposed in~\cite{DBLP:journals/is/ChapelaCampaBBDKS25} that have been extended with some new introduced metrics. ii) \textbf{Prediction quality preservation} was evaluated using the Train-on-Synthetic-Test-on-Real framework proposed in~\cite{Refaeilzadeh2009}, which involves training the same machine learning classifier separately on the generated and real datasets, and then comparing their performance on a shared real test set. iii) The \textbf{log variability} in the generated event-logs is analyzed using the concepts of trace and prefix entropy, as introduced in~\cite{DBLP:journals/jodsn/BackDS19}, as well as the notions of activity and trace duration, which are defined in this paper. Finally, an analysis of iv) \textbf{Computational times} taken by the different techniques is reported. 

After a brief summary comparing the results obtained by the different techniques, the most performant ones were further evaluated in a practical predictive process monitoring scenario. In this setting, the techniques were used to generate synthetic process traces containing activities that are rare in the original event logs, thereby increasing their frequency. A common classifier was then applied to each scenario to predict the occurrence of the target activity, and the results obtained by the various models were compared.

A comparison was also made with the Synthetic Minority Over-sampling Technique (SMOTE)~\cite{smote}: a widely used data augmentation method designed to address class imbalance. 
SMOTE does not consider the constraints on the different process perspectives when augmenting. The results indeed confirm that SMOTE does not help to improve accuracy in prediction, confirming the hypothesis that it is crucial to consider the process constraints when performing event-log augmentation.

The remainder of this paper is organized as follows: Section~\ref{sec:related_works} starts reporting on the state-of-the-art in augmentation/generation of datasets in AI, in general, and then concludes summarizing the literature focusing on augmentation/generation of event-logs. Section~\ref{sec:preliminaries} introduces preliminary concepts, including events, traces, and their characteristics. Section~\ref{sec:framework} briefly introduces the baseline technique, which will be integrated in the comparison with other techniques. Section~\ref{sec:experiments} reports on the evaluation framework, the metrics employed, and the results obtained, while in Section~\ref{sec:aug_for_accuracy}, the best techniques from Section~\ref{sec:experiments} were tested by generating synthetic traces with rare activities and comparing classifier performance. Section~\ref{sec:conclusion} provides a summary of the article's contribution. Along with this article are provided two appendices detailing the baseline technique introduced and its hyperparameter optimization. 

\section{Literature Analysis}\label{sec:related_works}
\noindent 
This section examines the key data augmentation techniques developed, from early statistical methods to modern generative adversarial networks, highlighting applications across various domains including process mining. Section~\ref{sec:related_works_augmentation} reports on the evolution of data augmentation techniques, while Section~\ref{subsec:aug_in_pm} reports on the various data augmentation techniques from the process mining state-of-the-art.

\subsection{Data Augmentation in AI}\label{sec:related_works_augmentation}
\noindent A large body of research has demonstrated the effectiveness of data augmentation in improving the performance of AI and machine/deep learning models.

Early techniques to synthetic data generation predominantly relied on statistical methods and simulation-based techniques. Rubin in~\cite{rubin2004multiple} introduced the concept of multiple imputations using chained equations (MICE) to handle missing data, which laid the groundwork for creating synthetic datasets by iteratively estimating missing values. Lately, Chawla et al.~\cite{smote} introduced the widely spread “Synthetic Minority Over-sampling Technique”(SMOTE), a popular method for addressing class imbalance in datasets. It generates synthetic samples by interpolating existing minority class samples, thus aiming to improve the performance of classifiers and predictors in imbalanced datasets.

The application of synthetic data spans multiple domains. In Natural Language Processing (NLP), Easy Data Augmentation~\cite{wei-zou-2019-eda} is a method that introduces random augmentations such as random insertion, deletion, and replacement of words. In computer vision, several augmentation methods have been proposed, ranging from basic pixel-level transformations to advanced strategies like ReMix~\cite{remix} and SmoothMix~\cite{smoothmix}, which improve model robustness by mixing samples in the input or feature space. In the time series domain, data augmentation has been surveyed in~\cite{survey_ts}.

More recent advances leverage neural methods to improve robustness and generalization. In computer vision, Generative Adversarial Networks (GANs)~\cite{goodfellow} have been widely adopted to generate high-fidelity data to augment training datasets. In the context of language, Large Language Models (LLMs) are now used to generate high-quality synthetic text data~\cite{DBLP:conf/iclr/GaoPLXY0ZLLK23,augpt}. These generative techniques not only expand the training distribution but also introduce controlled variation, leading to models that generalize better to unseen inputs, hence increasing performance.


\subsection{Data Augmentation in Process Mining}\label{subsec:aug_in_pm}
\noindent Despite the advances in AI and machine/deep learning (see Section~\ref{sec:related_works_augmentation}), process mining remains relatively underexplored in terms of data augmentation.
A large body of existing research on data augmentation relies on the assumption that training examples are independent of one another~\cite{data_survey}. Conversely, in event-log traces, events are inherently dependent on one another, for instance, process constraints may prohibit certain activities from occurring after specific preceding events. Therefore, traditional data augmentation techniques are not applicable in process mining and analytics, since they are all based on the assumption that the samples are stochastically independent. However, as outlined by Chapela-Campa et al.~\cite{dumasaug} and by Dumas et al.~\cite{ghidiniaalstaug}, the problem of data augmentation in process mining has recently gained momentum. 

The work by van Straten et al.~\cite{vanStratenPadellaHassani2025} introduced an augmentation framework for event logs. The framework is inspired by techniques from natural language processing, incorporating domain knowledge to augment the dimensionality of the event logs. However, this work only focuses on the generation of event logs that only indicate the activities performed, i.e. the control-flow. For this reason, we decided to not include this works in the comparison. Analogously for the work by Käppel and Jablonski~\cite{10.1007/978-3-031-34560-9_23}, which only focuses on control-flow augmentation, assigning placeholders for resource and timestamps. 

While recent research has started to address data augmentation in process mining, a substantial body of work has emerged focusing on the generation of synthetic event data. Although these techniques are not always explicitly designed to improve the performance of machine learning predictors, they share similar goals to enhance data availability and generalization. The work by Rozinat et al.~\cite{rozinat} is one of the first using process mining techniques to discover multiple perspectives of a process (control-flow, time, resources) and integrating them into complete simulation models. They can then be used for “what-if” analysis and operational support generating complete and realistic event-logs. 

In recent years, the increasing availability of data and the advancement of new machine and deep learning techniques led to the integration of these into traditional process simulation methods. Camargo et al.~\cite{DBLP:journals/corr/abs-2009-03567} investigated the benefits of employing deep learning techniques compared to traditional methods, finding that deep learning techniques produce more accurate synthetic data, although their technique does not enable “what-if” analyses. Consequently, new research is moving towards hybrid models, which combine traditional statistical and process mining techniques with machine and deep learning~\cite{CAMARGO2023102248,DBLP:journals/is/MeneghelloFGR25}. The major notable and recent contributions in the field of syntetic event-log generation include:
\begin{description}
    \item [\textbf{DSIM}] by Camargo et al.~\cite{CAMARGO2023102248}. It combines data-driven simulation and deep learning techniques to construct hybrid process simulation techniques. It begins by extracting a stochastic process model using the Split Miner algorithm~\cite{split_miner}, which captures the structure and control-flow of the process. The process model is subsequently enhanced with a generative deep learning component, employing a Long Short-Term Memory network to learn the temporal dynamics from the same event logs, incorporating resources' role information. This integration enables the technique to produce timestamped event sequences.
    \item [\textbf{LSTM}] by Camargo et al.~\cite{LSTM_camargo}. It generates sequential events by predicting subsequent occurrences based on prior data. The proposed technique initiates with an embedding layer, followed by two concatenated LSTM layers, which are crucial for capturing and learning the temporal dependencies and patterns inherent in the data, also associated with resources' role. Recursively using this LSTM-based predictive technique generates the next event in a sequence.
    \item [\textbf{LSTM (GAN)}] by Taymouri et al.~\cite{10.1007/978-3-030-58666-9_14}. It generates sequential events by predicting subsequent occurrences based on prior data. However, the training of the neural network was done following the Generative Adversarial Neural Network paradigm, where two LSTM neural networks are put against each other in a two-player game. The first is the Generator, that generates synthetic data, while the second, the Discriminator, tries to recognize whether a generated process instance is fake or not. Once the network has been trained, the Generator has been used as described in the work from~\cite{LSTM_camargo} for generating synthetic process instances.
    \item [\textbf{RIMS}] by Meneghello et al~\cite{DBLP:journals/is/MeneghelloFGR25}. This framework introduces a data-driven technique to generate process instances by integrating deep learning and Discrete Event Simulation within a white-box generation framework. It dynamically incorporates predictions based on Long Short-Term Memory techniques during the simulation, which mimics the procedure employed by DSIM. It facilitates inter-case feature calculations such as ongoing trace counts and the inclusion of real-time queue information. While it uses the same process discovery technique as the DSIM technique, it employs a distinct method for discovering branching probabilities.
    \item [\textbf{SIMOD}] by Chapela-Campa et al.~\cite{simod}. It presents a data-driven framework that automatically discovers and optimizes business process techniques from execution logs, and it is able to generate traces. It decomposes the problem into a series of steps with associated configuration parameters. A hyper-parameter optimization method is then used to search through the space of possible configurations to maximize the similarity between the behaviour of the generation technique and the behaviour observed in the log. Furthermore, the framework is able to classify data attributes and discover their associated update rules through an algorithm based on recursive update rules~\cite{DBLP:conf/icpm/LopezPintadoMD24}.
    \item [\textbf{AgentSimulator}] by Kirchdorfer et al.~\cite{DBLP:conf/icpm/KirchdorferBKAS24}. It presents a resource-centered technique for simulating event logs. It employs a multi-agent system derived from event logs, where resources are treated as autonomous agents interacting to simulate process execution. Within this framework, that remains white-box, the initial step involves defining the resources alongside a set of general post-hoc parameters. Subsequently, the simulation is executed by associating activities and timestamps with these defined resources.
    
    \item [\textbf{CVAE}] by Graziosi et al.~\cite{DBLP:conf/icpm/GraziosiRBFFGMP24,Graziosi2025}. It presents a method that has been solely used for data generation and relies on a Conditional Variational Autoencoder based on a LSTM network for generating new synthetic data. This work is not only capable of generating syntetic traces, but also explores the potential of Conditional Variational Autoencoder that offers control over the generation process by tuning input conditional variables, enabling more targeted and controlled data generation. According to Section~2 in~\cite{DBLP:conf/icpm/GraziosiRBFFGMP24,Graziosi2025}, In CVAEs, both the encoder and the decoder take the input data and conditioning variables as inputs. The conditional variable represents the specific condition or attribute that guides the generation process. However, as the aim of this paper is to compare various techniques without emphasizing the generation of a particular event or perspective, we opted to exclude any form of conditioning during the generation phase.
\end{description}

These techniques demonstrate a wide range of techniques tailored for different purposes, including “what-if” analysis, operational support, predictive process monitoring, and synthetic data generation. However, despite their potential, their capacity to generalize and their effectiveness for data augmentation tasks remain largely unexplored. In this paper, we conduct a comprehensive experimental comparison of state-of-the-art methods (see Section~\ref{sec:experiments}). However, since all existing approaches rely on machine or deep learning frameworks, we additionally introduce a novel baseline for data augmentation that depends solely on a stochastic transition system and probabilistic sampling (see Section~\ref{sec:framework}). Through this analysis, we aim to provide valuable insights into the applicability and limitations of current event data generation techniques.

\section{Preliminaries}\label{sec:preliminaries}\noindent
\noindent
Process data are typically collected in event logs. Event logs are collection of traces, each of which consisting of sequences of events~\cite{vanderAalst2016}. 
\begin{definition}[Events]
    Let $\mathcal{C}\subset\mathbb{N}$ be the set of trace identifiers. Let $\MA$ be the set of process activities. Let $\MR$ be the set of possible resources. Let $\MT\subset\mathbb{R}^{>0}$ the set of timestamps. Let $\MI$ be the set of possible event lifecycles. Let $\MV=\MV_1\times\ldots\times\MV_m$ be the cartesian product of the data attribute sets. An event is here defined as a tuple $(c, a, r, t, l, \vec{v})\in\mathcal{C}\times\MA\times\MR\times\MT\times\MI\times\MV$.
    
\end{definition}

Let $\ME$ be the universe of events. Given an event $e\in\ME$ we assume the following projections:
\begin{itemize}
    \item $case(e)\in\mathbb{N}$: the \textbf{trace identifier}, indicating which process case the event belongs,
    \item $act(e)\in\MA$: the \textbf{activity} being executed,
    \item $res(e)\in\MR$: the \textbf{resource} involved in executing the event,
    \item $time(e)\in\mathbb{R}^{>0}$: the \textbf{timestamp} of the event,
    \item $life(e)\in\MI$: the \textbf{lifecycle transition} of the event (e.g. \textit{start}, \textit{complete}, \textit{schedule}),
    \item $\mathit{attr}(e) \in \MV = \mathcal{V}_1 \times \ldots \times \mathcal{V}_m$: the vector of \textbf{event attributes}, where each $\mathcal{V}_i \in \{\mathcal{V}_1, \ldots, \mathcal{V}_m\}$ denotes the set of all possible values for the $i$-th attribute of the process. The component $\mathit{attr}(e).i \in \mathcal{V}_i$ represents the value of the $i$-th attribute for the event $e$.

\end{itemize}

In addition, resources are grouped on their specific \textbf{roles}. Specifically, a role includes a set of resources that can perform only a set of defined activities. We then introduce a further function $role:\ME\rightarrow\ 2^{\MR}$, that, given an event $e\in\ME$, returns a set of resources $\mathcal{R'}\in2^{\MR}$ that can perform the activity $act(e)$.

A trace is a sequence of events. The same event can occur in different traces. Namely, attributes may be given the same assignment in different traces. This means that the same trace can appear multiple times, although admittedly under extremely rare conditions, and motivates why an event log has to be defined as a multiset of traces:

\begin{definition}[Traces \& Event Logs]
    Let  $\ME=\mathcal{C}\times\MA\times\MR\times\MT\times\MI\times\MV$ be the universe of events. A trace $\sigma\in\ME^*$ is defined as a finite sequence of events ordered by timestamp and sharing the same case identifier. Specifically, $\sigma=\langle e_1,\dots,e_n\rangle\in\ME^*$ s.t. $case(e_i)=case(e_j), \forall i,j = 1,\dots,n$ and $time(e_i)\leq time(e_{i+1})\ \forall i=1,\dots,n-1$. An event log $\ML$ is a set of such traces: $\ML \subseteq \ME^*$.
\end{definition}

 To capture the different states in which a non-completed trace we use the \textbf{prefixes} of traces, i.e. sequence of events that represent the process execution up to a certain point.
Given a trace $\sigma=\left\langle e_{1}, \ldots, e_{n}\right\rangle$, the set of possible prefixes is defined as: 
$prefix(\sigma)=\left\{\langle\rangle,\left\langle e_{1}\right\rangle,\left\langle e_{1}, e_{2}\right\rangle, \ldots,\left\langle e_{1}, \ldots, e_{n}\right\rangle\right\}$.
However, using full-length prefixes may lead to overfitting and poor generalization. To address it, we introduce the notion of \textbf{$k$-prefixes}, that restricts the conditions n the most recent $k$ events.
For a length $k>0$, we define $prefix_k(\sigma)=\left\{\langle\rangle,\left\langle e_{1}\right\rangle,\left\langle e_{1}, e_{2}\right\rangle, \ldots,\left\langle e_{j}, \ldots, e_{j+k}\right\rangle,\dots, \langle e_{n-k},\dots,e_n\rangle\right\}$.
 
This formulation includes all contiguous subsequences of the trace with at most $k$ events. If $k \geq n$, then $prefix_k(\sigma) = prefix(\sigma)$. 
Also, we refer to \textbf{duration} of a trace, the difference within the highest and the lowest timestamp, i.e. $dur(\sigma)=time(e_n) - time(e_1)$. 
Analogously, we define the duration of an activity as follows:


\begin{definition}[Activity Duration]
Let $e \in \sigma$ be an event of a trace $\sigma \in \mathcal{L}$. 
The duration of $e$ is defined as:
\[
\resizebox{0.9\linewidth}{!}{$
dur(e) =
\begin{cases}
time(e') - time(e), & 
\begin{aligned}[t]
&\text{if } life(e) = start,\ \exists e' \in \sigma\ s.t.:\\
&act(e') = act(e)\ \land\  life(e') = complete
\end{aligned} \\[1.2em]
time(e) - time(e'), & 
\begin{aligned}[t]
&\text{if } life(e) = complete,\ \exists e' \in \sigma\ s.t.:\\
&act(e') = act(e)\ \land\ life(e') = start
\end{aligned}
\end{cases}
$}
\]
\end{definition}
Note that this definition represents a simplification, as multiple events with the same activity and lifecycle may occur within the same trace; in such cases, we consider the pair of events whose timestamps are closest in time.

The interaction of resources in a process is one of the main perspectives to be analyzed in process mining. A social network of resources can be built based on the interactions between resources using the concept of \textbf{handover of work}, that was introduced in~\cite{vanderAalst2016}.

\begin{definition}[Handover of Work]
    Let $\ML$ be an event log. Let $r_i,r_j\in\MR$ be two different resources. The number of times resource $r_i$ hands over work to resource $r_j$ is given by: \[\resizebox{\textwidth}{!}{$hw_{\ML}(r_i,r_j) = \sum_{\langle e_{1}, \ldots, e_{n}\rangle \in \ML}\mid \{ (e_k,e_{k+1}) : 1 \leq k \leq n-1 ,\; res(e_k)=r_i \land res(e_{k+1})=r_j \} \mid $} \]
\end{definition}

The quantity $hw_{\ML}(r_i, r_j)$ then expresses how often the resource $r_i$ passes work to the resource $r_j$, thus indicating how strong their relationship in the process. Note that this holds in one direction, since $hw_{\ML}(r_i, r_j)\neq hw_{\ML}(r_j, r_i)$.

Furthermore, the timestamps recorded in event logs allow for the modeling of temporal aspects of process execution. As reported in~\cite{DBLP:journals/is/ChapelaCampaBBDKS25}, the \textbf{inter-arrival time} refers to the time elapsed between the arrivals of two consecutive cases. while the \textbf{inter-execution time} of an event denotes the duration between that event and its immediate predecessor. These temporal dimensions are subsequently employed to capture the temporal perspective in our approach.

\section{A Baseline Technique for Event-Log Augmentation}\label{sec:framework}\noindent

\noindent Section~\ref{sec:related_works} illustrates a number of techniques that leverage on different deep- or machine-learning techniques. These techniques naturally requires a high computation workload. An interesting research questions is whether this workload is actually justified, if compared with simpler techniques. 
For this reason, we introduce a baseline technique based on a stochastic transition system, which is quick to learn and use, since it solely rely on statistical methods. This baseline still allows generating the process perspectives on control-flow, resource, time and data attributes.   




The baseline is not meant to be the main contribution of this article, and hence this section only describes it in a nutshell. Interested readers to gain full insight can refer to~\ref{appendix:probagen}, where a description of the different steps is provided. 

\begin{figure}[t!]
    \centering
    \includegraphics[width=\linewidth]{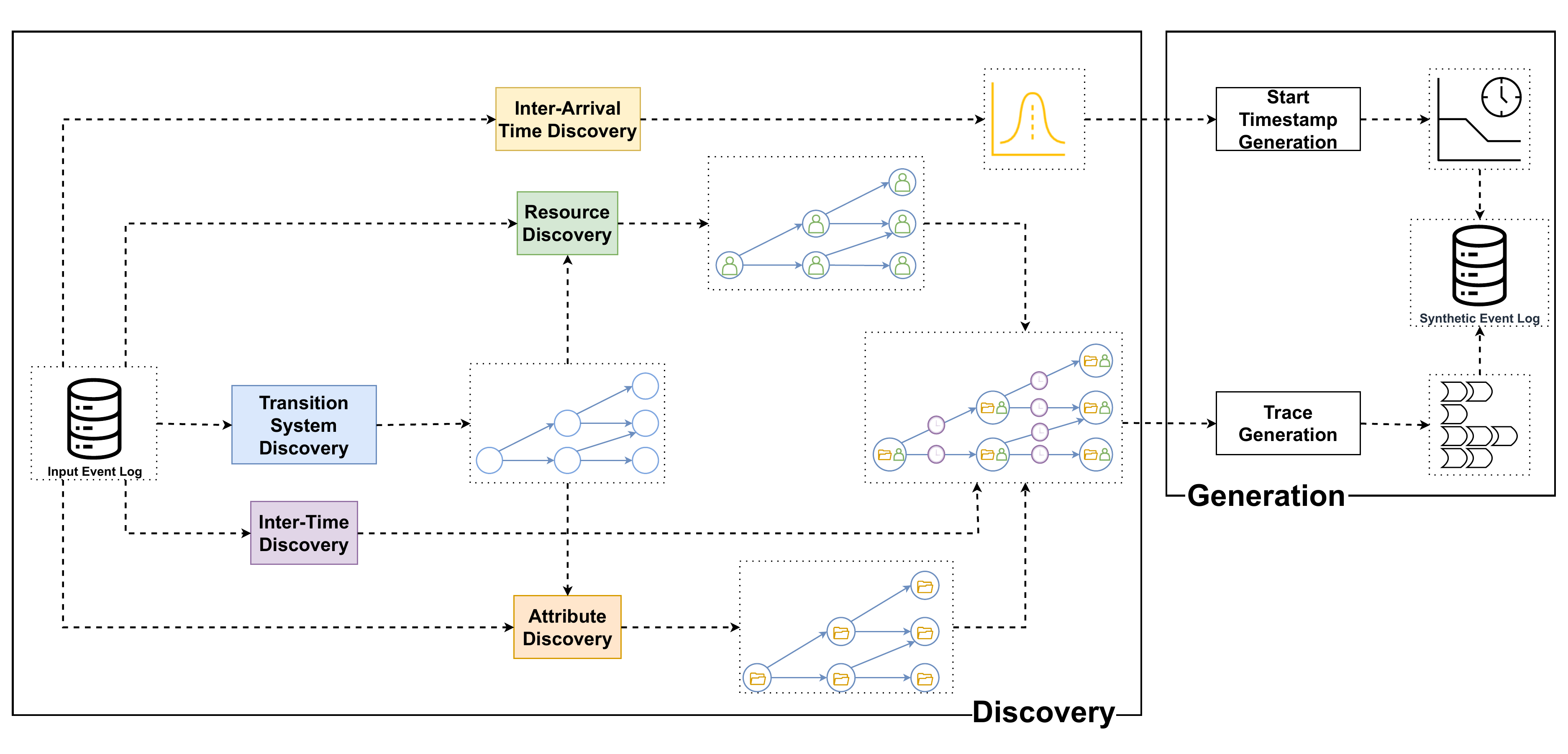}
    \caption{Overview of the Baseline technique. In a first \textit{Discovery} phase, an input event log is used for generating a stochastic transition system with various perspectives. Time distributions as derived from the event log. In the \textit{Generation} phase, the transition system is used for generating traces, while the inter-arrival distribution for sampling starting timestamp for each trace. These result in a synthetic event log.}
    \label{fig:proba_diagram}
\end{figure}

\autoref{fig:proba_diagram} illustrates the baseline technique, which is divided into two phases: \textit{Discovery} and \textit{Generation}. The starting point is an input event log $\ML$, from which a probabilistic transition system is constructed to capture the probabilistic behavior of the process. In this system, each state represents a trace prefix, and each transition corresponds to the execution of an activity, moving from one prefix to another. Transitions are associated with probabilities that indicate the likelihood of moving from one state to the next. Finally, this transition system is used to generate a synthetic event log during the Generation phase.

Here are highlighted the different parts of the \textit{Discovery} phase, with an associated example:
\begin{description}
    \item[Control-flow Discovery.] Given a prefix of activities, the system computes the probability over the possible next activities. For example, in a loan application process, after \textit{Submit Application}, the system may estimate that the next step is \textit{Check Credit Score} with a probability of 0.75, \textit{Cancel Application} with 0.15, or \textit{Request Additional Documents} with 0.10.
    \item[Resource Discovery.] For each activity occurrence, conditioned on its prefix and previously assigned resources, the system estimates who is likely to execute it. For example, if \textit{Officer A} handled \textit{Submit Application}, the probability that the same officer also executes \textit{Request Additional Documents} may be 0.8. Conversely, if \textit{Officer B} submitted the application, the probability that they handle the request step may only be 0.3, with the remainder split among other officers. This captures continuity and handover-of-work patterns.
    \item[Event Attributes Discovery.] Similarly, additional attributes are sampled on the basis of current activity and the history of previously assigned attributes. For example, if a case was submitted with attribute \textit{LoanType=Housing Loan}, then during \textit{Check Credit Score} the probability of assigning \textit{Priority=High} may increase to 0.6 (versus 0.4 for low priority).
    \item[Temporal Discovery.] Both inter-event times (the delay between two consecutive events in a trace) and inter-arrival times (the time between the start of two traces) are estimated via best-fitting distributions. For example, the inter-event time between \textit{Submit Application} and \textit{Check Credit Score} might follow a log-normal distribution with mean 2 hours. Inter-arrival times between cases may follow an Exponential distribution with an average of one new loan application every 15 minutes.
\end{description}
\noindent Finally, in the \textit{Generation} phase, trace arrival times are generated sampling from the inter-arrival time distribution. Then, for each trace, events are sequentially generated by randomly walking the stochastic transition system according to transition probabilities. At each step, the next activity is sampled based on the current prefix, and corresponding resources and attributes are sampled conditionally. Timestamps are assigned by sampling inter-event times from the appropriate distribution and adding them cumulatively. The result is a complete synthetic event log.

This technique has been implemented in Python and is publicly available online. It takes an event log as input and allows the generation of a new synthetic log with a user-defined number of traces.\footnote{\url{https://github.com/franvinci/ProbabilityBasedEventLogGenerator/tree/complete-generator}}

\section{Experimental Comparison}\label{sec:experiments}\noindent
This section reports on the experiments conducted to compare the quality of event-logs generated by the techniques reported in Section~\ref{sec:related_works} and by the baseline introduced in Section~\ref{sec:framework}. Different techniques were evaluated under four different points of view:
\begin{description}  
 
    \item [\textbf{Similarity}:] To ensure the coherence of the generated logs with respect to the real ones, a set of state-of-the-art distance metrics from~\cite{DBLP:journals/is/ChapelaCampaBBDKS25} has been employed, which have also been extended with two additional metrics for testing the quality of generation of roles and data attributes.

    \item [\textbf{Predictions quality preservation}:] The \textit{Train-on-Synthetic-Test-on-Real} (TSTR) method, introduced in~\cite{tstr} has been used to assess whether the predictive power of the generated logs is equivalent to that of the original logs.

    \item [\textbf{Log variability through entropy}:] Leveraging and further extending the concept of entropy of an event log introduced in~\cite{DBLP:journals/jodsn/BackDS19}, we measure the difference in information contained between the real and the generated event-logs. 

    \item [\textbf{Computational times:}] A comparison of computational times needed for both training the models and generating the logs has been provided.
   
\end{description}

The remainder of this section is organized as follows: Section~\ref{subsec:train_test_split} details the methodology for splitting the logs into training and testing datasets. Section~\ref{subsec:generation-metrics} describes the evaluation metrics used on the eight distinct case studies, outlined in Section~\ref{subsec:use_cases}. Finally, the Section~\ref{subsec:res_analysis} reports the analysis of the results.

\subsection{Evaluation Methodology}\label{subsec:train_test_split}
The similarity of the generated logs has been assessed from multiple perspectives.
To prevent information leakage, event logs have been split temporally into train and test sets, $\ML^{train}$ and $\ML^{test}$, with respectively 80\% and 20\% of the total traces, according to most of the techniques in this survey. Initially, $\ML_{train}$ is employed for generating a log $\ML_{gen}$ with the same number of traces as $\ML_{test}$, and then the quality of generated logs is assessed through the perspectives that will be introduced in Section~\ref{subsec:generation-metrics}. \textit{All the reported metrics are the average of the evaluation on 10 different generated logs for mitigating the effect of the stochasticity below some generative techniques.}



\subsection{Evaluation Metrics}\label{subsec:generation-metrics}\noindent

To assess the quality of the generated event logs, we employ a comprehensive set of evaluation metrics. These metrics analyze various aspects of the generated traces, including their similarity to the real event log, their ability to preserve or improve predictive performance when used for training a regressor, the information content measured through entropy, and the computational time required to generate the synthetic traces.

\subsubsection{Similarity of Generated Logs}\label{sec:metrics}\noindent
In this subsection, we present the metrics employed to evaluate the similarity between the generated log $\ML^{gen}$ and the real test log $\ML^{test}$, based on the comparison of their respective properties. A variety of metrics are used:

\begin{itemize}

    \item \textbf{Control-flow Log Distance} (CFLD), that given two logs $\ML_1$ and $\ML_2$, compute the average distance to transform each sequence of activities in $\ML_1$ into another in $\ML_2$.

    \item \textbf{N-Gram Distance} evaluates the difference in the frequency distribution of activity sequences between two logs. To be consistent with the paper~\cite{10.1007/978-3-030-49435-3_14} that introduces the distance, the technique has been tested for the 2-Grams, for which it is equivalent to the Earth Mover Distance (EMD)~\cite{e19020047} evaluated in~\cite{DBLP:journals/corr/abs-2009-03567}, and 3-Grams.

    \item \textbf{Absolute Event Distribution} (AED) Given two logs $\ML_1$ and $\ML_2$, the timestamps associated with a trace are transformed into time series, where at each timestamp correspond a number of events, and then the EMD between the series is evaluated as in~\cite{710701}.

    \item \textbf{Relative Event Distribution} (RED) evaluates how accurate is the generation of the timestamp of subsequent events based on preceding events. For each case a timeseries of number of events within that case is built, and then the EMD is computed.

    \item \textbf{Circadian Event Distribution} (CED) Given two logs $\ML_1$ and $\ML_2$, they are divided into subsets of events by day of the week and then the EMD is computed.

    \item \textbf{Cycle Time Distribution} (CTD) That measures the EMD between the distribution of the total cycle times of the traces in the event-logs.
    
    \item \textbf{Case Arrival Rate} (CAR) that compares the distribution of the inter-arrival times of two event-logs.

    \item \textbf{Circadian Workforce Distribution} (CWD) that computes the average difference in the distribution of active resources throughout the process timeline, by day of the week.

\end{itemize}

\noindent
The reported set of metrics, introduced in~\cite{DBLP:journals/is/ChapelaCampaBBDKS25}, covers several key perspectives of event logs, however, existing techniques lack dedicated metrics for analyzing \textit{roles}, \textit{handover}, and \textit{attributes} generation.

To bridge this gap, we introduce a set of complementary metrics. The first among them is the \textbf{Role-Based Circadian Event Distance (RCED)} is introduced to assess the quality of the generated roles of the resources (cf. Section~\ref{sec:preliminaries}). The Role-Based Circadian Event Distribution quantifies the similarity in the assignment of roles to activities between two event logs referring to the same process.

\begin{definition}\textbf{Role-Based Circadian Event Distribution (RBCED)} 
Let $\ML_1$ and $\ML_2$ be two event-logs. Let $role:\ME\rightarrow2^{R}$ be the role function and let $R\in cod(role)$ be a role. Let  
\[S(\mathcal{L}, R) = \bigcup_{\sigma \in \mathcal{L}} \bigoplus_{\substack{e \in \sigma \\ \mathrm{role}(e) = R}}e\]  Let CED be the Circadian Event Distance between two event logs. The RBCED is defined as the mean of the Circadian Event Distance for each role, namely \newline $RBCED(\ML_{1}, \ML_{2})=avg_{R\in cod(role)}CED(S(\ML_1,R),S(\ML_2,R))$.~\footnote{Considering $\displaystyle\bigoplus$ as the concatenation of vectors e.g. $[1,3,'request\_ created']\bigoplus[2,True]=[1,3,'request\_ created',2,True]$}
\end{definition}

Note that while DSIM, RIMS and LSTM focus on associating roles to events, while SIMOD, CVAE, AgentSimlator and the baseline associate resources. This makes the RBCED unfeasible for some of these techniques. To solve this issue, roles have been inferred using the role discovery technique proposed by~\cite{6597224}. In it, each activity is associated with a different role and is associated with the multiset of his originators. After that, roles are merged according to their similarity until no more merges are possible associating each activity with a role. Then, roles are assigned to resources based on the most frequent role they have in the generated log.

\noindent Then, we introduce the Handover-of-Work Distance (HWD) to capture differences between event-logs from a resource interaction perspective (i.e. the handover of work, introduced in Section~\ref{sec:preliminaries}). 
The following defined distance quantifies how much the handover behavior between resources differs across two event-logs.
\begin{definition}\textbf{Handover-of-Work Distance (HWD)}
    Given two event-logs $\ML_1$ and $\ML_2$, with a set of involved resources $\MR=\{res(e)|e\in\ML_1\cup\ML_2\}$, the Handover of Work Distance between $\ML_1$ and $\ML_2$ is defined as: 
    \[
    HWD(\ML_1,\ML_2)=\sum_{r_i,r_j\in\MR}\left| hw_{\ML_1}(r_i,r_j)-hw_{\ML_2}(r_i,r_j)\right|
    \]
\end{definition}
Finally, we introduce the \textbf{Data Attribute Distribution (DAD)} that computes the EMD between the distributions of data attributes in two event logs.

\begin{definition}\textbf{Data Attribute Distribution (DAD)}
Let $\ML_1$ and $\ML_2$ be two event logs, and let $\mathcal{V}$ denote a set of attributes. The Data Attribute Distribution is then computed as the average EMD between the observed distributions of attributes in the two event logs $\ML_1$ and $\ML_2$:\footnote{Symbol $\uplus$ indicates the union of sets, generating a multiset.}

\[
DAD(\ML_1,\ML_2)=avg_{i=1,\ldots,m}EMD(V^i_{\ML_1},V^i_{\ML_1})
\]


\[
V^i_{\ML_1}=\biguplus_{\sigma\in\ML_1} \{attr(e).i \mid e\in\sigma\}, \quad  
V^i_{\ML_2}=\biguplus_{\sigma\in\ML_2} \{attr(e).i \mid e\in\sigma\}.
\]

\end{definition}

\noindent
All the distance measures introduced in this section are associated with specific perspectives of the event log. In particular, the CFLD and N-Gram metrics capture aspects of the \textit{control-flow} perspective, whereas AED, RED, and CED pertain to the \textit{temporal} perspective. The \textit{congestion} perspective is characterized by the CTD and CAR metrics. Furthermore, RBCED, CWD, and HWD describe the \textit{resource} perspective, while DAD reflects the \textit{attribute} perspective. A summary of these relationships is provided in Table~\ref{tab:metric_sumup}.
This grouping follows the same rationale as~\cite{DBLP:journals/is/ChapelaCampaBBDKS25}, with the addition of the DAD metric for measuring the similarity of attributes generation, and the addition of two new resource metrics (RBCED and HWD), which are grouped with CWD.

\begin{table}[h]
\centering
\begin{tabular}{ll}
\toprule
\textbf{Perspective} & \textbf{Associated Distances} \\
\midrule
Control-flow & CFLD, N-Gram \\
Temporal & AED, RED, CED \\
Congestion & CTD, CAR \\
Resource & RBCED, CWD, HWD \\
Attribute & DAD \\
\bottomrule
\end{tabular}
\caption{Summary of distance metrics and their corresponding event log perspectives.}
\label{tab:metric_sumup}
\end{table}

\subsubsection{Prediction Quality Preservation}\label{tstr_metric}\noindent
The \textit{Train-on-Synthetic-Test-on-Real} (TSTR) approach is employed to assess how well predictive performance is preserved when models are trained on the generated synthetic data~\cite{tstr,10.5555/3454287.3454781,VALLEVIK2024105413}.
Let $\ML_{train}$ and $\ML_{test}$ denote the training and test logs, respectively, as defined in Section~\ref{subsec:train_test_split}. In the first stage of the application of this technique, the user initially defines a dependent variable, and then two predictive models $\Phi_{train}$ and $\Phi_{gen}$ are trained on $\ML^{train}$ and $\ML^{gen}$, respectively, to predict it. In this paper, the defined dependent variable is the \textbf{Total Execution Time} of an individual trace. Predictors' performance is reported in terms of the \textbf{Relative Mean Absolute Error} (rMAE), defined as the Mean Absolute Error between the actual and predicted values, divided by the mean of the real values. This rescaling allows for a fair comparison across different logs, as the average trace durations vary substantially.

The models are evaluated on $\mathcal{L}^{run}$, which is obtained from $\mathcal{L}^{test}$ by truncating each trace to a randomly selected prefix of length $p$.\footnote{The prefix length percentage $p$ is sampled from a uniform distribution $\unif[25,\ 75]$, as in~\cite{10.1007/978-3-032-02936-2_15,bib:padella}. We remind that the reported results represent the average over 10 independent runs to account for the stochastic variability introduced by the sampling process.}
This procedure simulates running, i.e. non-completed, traces, while $\mathcal{L}^{test}$ provides the corresponding ground truth for evaluation. The predictive framework adopted follows the standard approach commonly used in predictive process monitoring (cf.~\cite{bib:padella}) and is implemented using the latest version of \textit{CatBoost}~\cite{Catboost}, an open-source, high-performance gradient boosting framework based on decision trees.\footnote{The \textit{CatBoost} reference implementation is available at \url{https://catboost.ai/}}

\subsubsection{Log Variability through Entropy}\label{subsubsec:entropy_measures}\noindent
The aim of event-log augmentation is not only to generate traces and events that are coherent with the real ones, but also to generalize the information contained in the event log by creating traces that are not present in the original event log. In information theory, this is measured through \textbf{entropy}~\cite{Jianbo2013Information}. Given a multiset $X$ of elements, where each element $x \in X$ is associated with a cardinality $p(x)$, the $X$'s entropy is computed as $H(X) = - \sum_{x \in \mathcal{X}} p(x) \log p(x)$. This definition originally given by Shannon is meaningfully applicable when $X$ includes values extracted from a domain that is not inherently discrete, such as when it is continuous or has a very high-resolution; otherwise, $p(x)$ tends to 0 for any $x \in X$, and therefore entropy tends to infinity.
To address this issue, we introduce the Discretized Entropy $\tilde H(X)$ as follows:

\begin{definition}[Discretized Entropy]
    Let  $X$ be a multiset of elements, partitioned in K disjoint intervals.
    Given a multi set $X$ and a partition of it into $K$ disjoint intervals (buckets) $\mathfrak{B}=\{B_1, B_2, \ldots, B_K\}$, we define the Discretized Entropy $\tilde H(X)$ as:
    
    \[
    \tilde H_{\mathfrak{B}}(X) = -\sum_{k=1}^{K} p_k \log (p_k)
    \]
    where $p_k = \frac{|B_k|}{|X|}$.
    
\end{definition}

To obtain the buckets, the data has been discretized using a bucketing technique that leverages and histogram for which the number of bins has been chosen using the technique introduced in~\cite{Scott2011}.

The concept of entropy has been used in process mining to measure how diverse the process executions are within a single event log. For this reason, we decide to measure the entropy of an event log following the guidelines introduced in~\cite{DBLP:journals/jodsn/BackDS19}. The first measure used is the so-called \textit{Trace Entropy}.

\begin{definition}[Trace Entropy]\label{def:trace_entropy}
    Let $\ML$ be an event log. Let $p$ be the frequency function. The trace entropy $H_{tr}$ of $\ML$, is defined as:

    \[ H_{tr}(\ML)=-\sum_{\sigma\in\ML}p(\sigma) \log p(\sigma)\]
\end{definition}

Similarly, we define the \textbf{Activity Duration Entropy}.

\begin{definition}[Prefix Entropy]
    Let $\ML$ be an event log. Let $p$ be the frequency function. The prefix entropy $H_{pr}$ of $\ML$, is defined as:

    \[ H_{pr}(\ML)=-\sum_{\substack{\sigma_i\in prefix(\sigma) \\ \sigma\in\ML}} p(\sigma_i) \log p(\sigma_i)\]
\end{definition}

The concept of entropy of traces and prefixes was introduced in~\cite{DBLP:journals/jodsn/BackDS19} and was only related to the control-flow perspective.
This is only one the process perspectives that we aim to generate and compare. For this reason, we introduce two additional entropy measures related to the time perspective.

\begin{definition}[Cycle Time Distribution Entropy]
    Let $\ML$ be an event log. Let $\tilde H_{\mathfrak{B}}$ be the Discretized entropy. Let $\mathcal{D}(\ML)=\displaystyle\biguplus_{\sigma\in\ML}\{dur(\sigma)\}$ be the set containing all the activity durations of the traces of $\ML$. The Cycle Time Distribution Entropy is defined as:
    \[H_{ct}(\ML)=\tilde H_{\mathfrak{B}}(\mathcal{D}(\ML))\]
\end{definition}

Similarly, we define the \textit{Prefix Entropy}.

\begin{definition}[Activity Duration Entropy]
    Let $\ML$ be an event log. Let $\tilde H_{\mathfrak{B}}$ be the Discretized entropy. Let $a\in\MA$ be an activity. Let $\mathcal{D}_a(\ML)=\displaystyle\uplus_{\sigma\in\ML}\{dur(e)|\ act(e)=a\}$. The Activity Duration Entropy $H_{ad}$ is defined as:
    \[
    H_{ad}=avg_{a\in\mathcal{A}}(\tilde H_{\mathfrak{B}}(D_a))
    \]
\end{definition}



We define the Activity Duration Entropy as the average of the entropy values computed individually for each activity, rather than evaluating entropy globally over all durations. This design choice aims to penalize low variability in the duration assignments of specific activities. For instance, if a generation technique produces constant durations for each individual activity, $H_{ad}$ will yield a value of zero. In contrast, computing entropy over the aggregated set of all activity durations without distinguishing between activity types may still result in a non-zero value, even when per-activity variability is absent.

\subsubsection{Workstation Specification}\noindent
All experiments were conducted on a dedicated workstation running Ubuntu 24.04.2 LTS (64-bit) with the Linux kernel version 6.8.0-62-generic. The machine was equipped with a 12th Generation Intel Core i7-12700KF processor featuring 20 cores, and supported by 32 GB of RAM. Graphics computations were accelerated using an NVIDIA GeForce RTX 4060 GPU, enabling efficient parallel processing during model training and inference.

The reported execution times reflect the total duration required to train the generator (if applicable) and to sample 10 different generations per event-log. This measurement technique was adopted to account for the variability in computational demands among the techniques under evaluation-some techniques incur significant training times but offer fast sampling, while others require minimal training times but longer sampling durations. This unified metric provides a fair basis for comparing the overall runtime efficiency across methods.

\subsection{Use Cases and Event Data Set}\label{subsec:use_cases}\noindent
The validity of the techniques was evaluated using eight different processes from which an event-log was extracted.

\begin{description}
    \item [BPI17:] The subprocess for the workflow-relevant in the 2017 BPI Challenge event data, a log of a loan application process from a Dutch financial institution. An event-log from the institution’s information system with 481,708 events and 30,276 completed traces has been extracted, divided into 24,221 for the train and 6,055 for the testing. Furthermore, the log contains 8 different activities that can be executed by 148 different resources.\footnote{\url{https://data.4tu.nl/articles/dataset/BPI_Challenge_2017/12696884}}

    \item [BPI12:] The log used by the BPI challenge in 2012\footnote{\url{https://doi.org/10.4121/uuid:3926db30-f712-4394-aebc-75976070e91f}}, and it is provided by the same financial institution that provides the log employed in \emph{BPI17}. 8,616 completed traces and 118,604 events have been extraced, divided into 6,892 for the train and 1,723 for the testing. It contains 13 different activities that can be performed by 58 different resources.

    \item [Production:] A process dealing with a manufacturing production company (MP). It has been exported from an Enterprise Resource Planning (ERP) system~\cite{https://doi.org/10.4121/uuid:68726926-5ac5-4fab-b873-ee76ea412399}. An event-log has been extracted from this process, containing 9,906 events and 225 completed traces, divided into 180 for the train and 45 for the test. Furthermore, the log contains 26 different activities that can be executed by 48 different resources.

    \item [Purchasing:] A process provided as part of the Fluxicon Disco tool and it is related to a purchase-to-pay (P2P) system, which is a synthetic and generated from a model not available to the authors\footnote{\url{https://fluxicon.com/academic/material}}. The extracted event-log has 18,238 events, recorded in 608 traces, divided in 486 for train and 121 for test. It contains 21 different activities that can be accomplished by 27 resources.

    \item [Consulta:] A process from an Academic Credentials Recognition (ACR) process of a Colombian University was gathered from its BPM system (Bizagi) for the fifth use case. The extracted event-log has 13,740 events, recorded in 954 traces, divided in 763 for train and 190 for test. It contains 18 different activities that can be accomplished by 561 resources.\footnote{\url{https://zenodo.org/records/5734443}}
    
    \item [Sepsis:] A log recording patient pathways with suspected sepsis, a life-threatening infection, over one year in a hospital. The extracted event-log has 15,214 events, recorded in 1050 traces, divided in 735 for train and 315 for test. It contains 16 activities that can be accomplished by 26 resources. \footnote{\url{https://data.4tu.nl/datasets/33632f3c-5c48-40cf-8d8f-2db57f5a6ce7}}
    
    \item [RTF:] Acronym for road traffic fines, is a process dealing with road fines that comes from an Italian local police information system handling traffic fines. The extracted event-log has 561,470 events, recorded in 150,370 traces, divided in 105,259 for train and 45,111 for test. It contains 11 activities that can be accomplished by 148 resources.\footnote{\url{https://data.4tu.nl/datasets/806acd1a-2bf2-4e39-be21-69b8cad10909}}

    \item[BPI19:] The log used by BPI challenge in 2019. It comes from a multinational coatings and paints company in the Netherlands, describing the purchase order handling process for its 60 subsidiaries. From this log, all the cases were evaluated only covering over 2.5 months of data from January to March 2018, in accordance with~\cite{DBLP:conf/icpm/LopezPintadoMD24}. This excerpt of the log contains 63,839 events, recorded in 9,743 traces, divided in 6,820 for train and 2,922 for test. It contains 34 activities that can be accomplished by 235 resources.
\end{description}

\noindent The first five processes were used to evaluate the quality of the generated measures, as all frameworks could be applied to these logs due to the presence of both starting and ending timestamps for each activity. In fact, in the BPI19, Sepsis and RTF only report the ending timestamps of the activities, making DSIM, LSTM, LSTM(GAN), RIMS, and AgentSimulator not applicable. 
Furthermore, the CWD metric cannot always be evaluated because LSTM, LSTM (GAN), and RIMS directly generate roles instead of resources, while DSIM only generates them synthetically with different names, making unfeasible the comparison with the real ones.

\subsection{Experimental Results}\label{subsec:res_analysis}
This section discusses the various experimental results, using the techniques introduced in Section~\ref{subsec:generation-metrics}. For every technique proposed, a hyperparameter optimization has been carried on following the procedure related to the corresponding work, while for the baseline approach, the hyperparameter optimization procedure is reported in~\ref{appendix:probagen_k}.
Section~\ref{subsec:similarity} discusses the level of similarity between the generated logs and the corresponding original ones. Section~\ref{pred} reports on the predictive performance of the generated logs, while Section~\ref{subsubsec:entropy_res} outlines the results in terms of entropy evaluation, and the computational times required for the various techniques are reported in Section~\ref{times}.
Finally, the results have been summed up in Section~\ref{sec:discussion}.

\subsubsection{Similarity of Generated and Real Event-Logs}\label{subsec:similarity}\noindent
The similarity results, computed on the test set as defined in Section~\ref{subsec:train_test_split}, for the various case studies are presented in Table~\ref{tab:results_distances_cfecc}. In addition, the final rows report the average performance of each technique for its respective metric. The rows represent the application of the different techniques discussed in Section~\ref{sec:related_works}, along with the baseline technique introduced in Section~\ref{sec:framework}. 

The columns of the table represent the evaluation metrics introduced in Section~\ref{sec:metrics}, grouped according to the perspective they capture, also reported in Table~\ref{tab:metric_sumup}: control-flow (CFLD, 2-Gram, 3-Gram) in green, temporal (AED, RED, CED) in red, congestion (CTD, CAR) in pink, and resource-related (RBCED, CWD, HWD) in yellow. 
As reported in Section~\ref{subsec:use_cases}, results for use cases RTF, Sepsis and BPI19 are not reported since the processes do not report the start timestamp of the activities, making most of the techniques not applicable, while RBCED and CWD are not reported for DSIM, LSTM, LSTM (GAN) and RIMS because they only generate roles, instead of resources. Since measures refer to distances,\textit{ lower values indicate better performance}.
\begin{table}[ht!]
    \centering
    \resizebox{\linewidth}{!}{\footnotesize
     \begin{tabular}{|l||>{\raggedright\arraybackslash}p{0.15\linewidth}|>{\columncolor{green!20}\raggedright\arraybackslash}p{0.07\linewidth}|>{\columncolor{green!20}\raggedright\arraybackslash}p{0.08\linewidth}|>{\columncolor{green!20}\raggedright\arraybackslash}p{0.08\linewidth}|>{\columncolor{red!20}\raggedright\arraybackslash}p{0.1\linewidth}|>{\columncolor{red!20}\raggedright\arraybackslash}p{0.1\linewidth}|>{\columncolor{red!20}\raggedright\arraybackslash}p{0.07\linewidth}|>{\columncolor{pink!20}\raggedright\arraybackslash}p{0.08\linewidth}|>{\columncolor{pink!20}\raggedright\arraybackslash}p{0.08\linewidth}|>{\columncolor{yellow!20}\raggedright\arraybackslash}p{0.09\linewidth}|>{\columncolor{yellow!20}\raggedright\arraybackslash}p{0.08\linewidth}|>{\columncolor{yellow!20}\raggedright\arraybackslash}p{0.08\linewidth}|}
     \hline
            \multicolumn{2}{|c|}{\textbf{\textit{Metrics' Perspective}}}    & \multicolumn{3}{c|}{\textbf{Control-flow}} 
          & \multicolumn{3}{c|}{\textbf{Temporal}} 
          & \multicolumn{2}{c|}{\textbf{Congestion}} 
          & \multicolumn{3}{c|}{\textbf{Resource}} \\
        \hline
        \hline
        Use Case & Model & \cellcolor{white} CFLD  &\cellcolor{white} 2-Gram   &\cellcolor{white} 3-Gram &\cellcolor{white} AED  &\cellcolor{white} RED   &\cellcolor{white} CED    &\cellcolor{white} CTD   &\cellcolor{white} CAR &\cellcolor{white} RBCED   &\cellcolor{white} CWD    &\cellcolor{white} HWD \\
        &\cellcolor{white} &\cellcolor{white} &\cellcolor{white} &\cellcolor{white} &\cellcolor{white} &\cellcolor{white} &\cellcolor{white} &\cellcolor{white} &\cellcolor{white} &\cellcolor{white} &\cellcolor{white}&\cellcolor{white} \\
        \hline
        \hline
     \multirow{8}{*}{\centering BPI17}
                & Baseline & 0.23 & 0.47 & 0.49 & 92.6 & 130 & \textbf{1.22}  & 124 & \textbf{35.9}  & 8.74 & 1.32 & 180426 \\
     \cline{2-13}& DSIM & 0.39 & 0.71 & 0.63 & 3401 & 110 & 2.44  & 142 & 3358  & 8.25 & - & -\\
     \cline{2-13}& LSTM & 0.39 & 0.84 & 0.61 & 8023 & 126 & 3.52  & 172 & 7934 & 19.3 & - & -\\
     \cline{2-13}& LSTM (GAN) & 0.64 & 0.93 & 0.88 &  192 & 40365 & 11.6  & 248 & 40332 & 25.8 &-& -\\
     \cline{2-13}& RIMS & 0.41 & 0.46 & 0.47 & \textbf{66.9} & 67.4  & 4.48  & 82.7 & 97.6 & 15.4 & - & -\\
     \cline{2-13}& SIMOD & 0.41 & 0.53 & 0.61 & 990 &  139 & 1.46 & 157 & 984 & 8.31 & \textbf{1.11} & 107919 \\
     \cline{2-13}& AgentSimulator & 0.41 & 0.61 & 0.69 & 89.2 & \textbf{19.9} & 1.44 & 67.6 &  103 & 9.88 & 1.72 & \textbf{51486} \\
     \cline{2-13}& CVAE & \textbf{0.19} & \textbf{0.13} & \textbf{0.16} & 2340 & 39.8 & 3.42 & \textbf{67.2} & 2421  & \textbf{8.02} & 3.60 & 100009 \\
    \hline
    \hline
     \multirow{8}{*}{\centering BPI12}                 & Baseline & 0.34 & 0.46 & 0.44 & 55.94 & 128  & 2.02  & 155 & 142 & 8.38 & \textbf{1.74}& 65850 \\
    \cline{2-13}& DSIM & 0.36 & 0.71 & 0.67 & 1630 & 191 & 5.69 & 172 & 1827 & 16.2 & - & -\\
    \cline{2-13}& LSTM & 0.18 & 0.53 & 0.45 & 1491& 107 & 7.73 & 106 & 1600  & 12.3 & - & -\\
    \cline{2-13}& LSTM (GAN) & 0.64 & 0.90 & 0.85 & 237 & 80475 & 17.9 & 206 & 80229 & 15.3 & - & -\\
    \cline{2-13}& RIMS & 0.34 & 0.51 & 0.62 & \textbf{31.7} & 71.8 & 3.42 & 84.3 & \textbf{78.7} & 14.6 & - & - \\
    \cline{2-13}& SIMOD & 0.25 & 0.45 & 0.49 & 225 & 114 & 1.90 & 103 & 2160 & 9.69 & 2.16 & 204255 \\
    \cline{2-13}& AgentSimulator & 0.40 & 0.65 & 0.66 & 56.9 & 154 & \textbf{1.36} & 199 & 134 & 7.82 & 1.88 & \textbf{10774} \\
    \cline{2-13}& CVAE & \textbf{0.14} & \textbf{0.12} & \textbf{0.13} & 930 & \textbf{14.2} & 4.27 & \textbf{13.3} & 931 & \textbf{6.57} & 4.18 & 22702 \\
    \hline
    \hline
    \multirow{8}{*}{\centering Production}
                & Baseline & \textbf{0.57} & \textbf{0.55} & \textbf{0.57} & \textbf{93.3} & 91.6 & 2.29  & 118 & \textbf{46.8} & \textbf{11.1} & \textbf{1.09} & 8431 \\
    \cline{2-13}& DSIM & 0.79 & 1.00 & 0.94 & 378 & 227 & 2.52 & 301 & 443 & 14.2 & -& -\\
    \cline{2-13}& LSTM & 0.87 & 1.00 & 1.00 & 389 & 275 & 18.0 & 360 & 441 & 15.2 & - & -\\
    \cline{2-13}& LSTM (GAN) & 0.91 & 1.00 & 1.00 & 78504 & 232 & 6.72 & 313 & 78575 & 11.4&-& -\\
    \cline{2-13}& RIMS & 0.80 & 0.80 & 0.81 & 103 &  \textbf{11.9} & \textbf{2.36} & \textbf{18.0} & 163 & 18.6 & - & --- \\
    \cline{2-13}& SIMOD & 0. 65 & 0.69 & 0.72 & 530 & 285 & 3.31 & 320 & 351 & 15.5 & 2.98 & 10283 \\
    \cline{2-13}& AgentSimulator & 0.75 & 0.95 & 0.96 & 1068 & 235 & 9.86 & 305 & 691 & 14.5 & 2.74 & 399 \\
    \cline{2-13}& CVAE & 0.77 & 0.85 & 0.92 & 480 & 124 & 2.21 & 283 & 310 & 16.5 & 1.93 & \textbf{371} \\
\hline
\hline
    \multirow{8}{*}{\centering Purchasing}& Baseline & 0.33 & \textbf{0.22} & \textbf{0.23} & 892 & 561 & \textbf{0.67}  & 434 & 652 & 7.2 & 0.81& 14945 \\
    \cline{2-13}& DSIM & \textbf{0.16} & 0.36 & 0.30 & 1134 &  722 & 1.00 & 596 & 774 & 11.2 & - & - \\
    \cline{2-13}& LSTM & 0.49 & 0.83 & 0.76 & 1300& 826 & 2.53 & 698 & 847 & 7.55 & - & -\\
    \cline{2-13}& LSTM (GAN) & 0.86 & 0.99 & 0.99 & 83375 & 782 & 3.56 & 638 & 83832 & 32.2 &-& -\\
    \cline{2-13}& RIMS & 0.79 & 0.81 & 0.89 & \textbf{103} & \textbf{12.3} & 2.13  & \textbf{17.5} & \textbf{157} & 17.9 & - & - \\
    \cline{2-13}& SIMOD & 0.54 & 0.97 & 0.98 & 79222 & 79435 & 4.13 & 62797 & 743 & 8.49 & 4.23 & 1611 \\
    \cline{2-13}& AgentSimulator & 0.75 & 0.90 & 0.95 & 1068 & 691 & 0.77 & 551 & 749 & 6.06 & 0.92 & \textbf{1207} \\
    \cline{2-13}& CVAE & 0.46 & 0.30 & 0.35 & 682 & 606 & 0.75 & 568 & 542 & \textbf{4.83} & \textbf{0.68} & 3944 \\
\hline
\hline  
    \multirow{8}{*}{\centering Consulta}& Baseline & 0.46 & 0.46 & 0.56 & \textbf{188} & \textbf{25.2} & 2.42  & 63.9 & \textbf{171} &6.52 & 2.91& 14209 \\
    \cline{2-13}& DSIM & 0.43 & 1.00 & 0.78 & 251 & 28.7 & 3.08 & 69.5 & 231 & 9.99 & - & -\\
    \cline{2-13}& LSTM & 0.44 & 1.00 & 0.79 & 543 & 40.0 & 21.1 & 106 & 503.8 & 19.2 &  - & -\\
    \cline{2-13}& LSTM (GAN) & 0.86 & 1.00 & 0.89 & 41162 & 37.9 & 14.3 & 85.2 & 41186 & 5.70 &-& -\\
    \cline{2-13}& RIMS & 0.56 & 0.64 & 0.80 & 242 & 31.4 & 3.12 & \textbf{61.0} & 234 & \textbf{6.2} & - & -\\
    \cline{2-13}& SIMOD & 0.36 & 0.36 & 0.39 & 898 & 502 & \textbf{1.81} & 504 & 227 & 8.16 & \textbf{1.58} & 79538 \\
    \cline{2-13}& AgentSimulator & 0.60 & 0.71 & 0.71 & 274 & 38.1 & 7.32 & 100 & 243 & 6.64 & 6.44& \textbf{1890} \\
    \cline{2-13}& CVAE & \textbf{0.30} & \textbf{0.30} & \textbf{0.33} & 691 &  66.5 & 4.74 & 134 & 587 & 10.6 & 4.62 & 3413 \\
\hline
\hline

\multirow{8}{*}{\centering \textbf{Average}}& Baseline & 0.39 & 0.43 & 0.45 & 264 & 187 & \textbf{1.72} & 178 & 209 & \textbf{8.39} & \textbf{1.57} & 56772 \\
\cline{2-13}& DSIM & 0.43 & 0.76 & 0.66 & 1358 & 255 & 2.95 & 256 & 1326 & 12.0 & - & -\\
\cline{2-13}& LSTM & 0.47 & 0.84 & 0.72 & 2349 & 274 & 10.5 & 288 & 2265 & 14.7 & - & -\\
\cline{2-13}& LSTM (GAN) & 0.78 & 0.93  & 0.96 & 40694 & 24432 & 10.8 & 298 & 64884 & 18.1 &-& -\\
\cline{2-13}& RIMS & 0.58 & 0.64 & 0.72 & \textbf{110} & \textbf{39.1} & 3.13 & \textbf{52.6} & \textbf{146.5} & 14.45 & -& - \\
\cline{2-13}& SIMOD & 0.44 & 0.60 & 0.64 & 16373 & 16095 & 2.52 & 12776 & 893 & 10.0 & 2.41 & 80721\\
\cline{2-13}& AgentSimulator & 0.58 & 0.76 & 0.79 & 511 & 227 & 4.15 & 244 & 384 & 8.99 & 2.74 & \textbf{13151} \\
\cline{2-13}& CVAE & \textbf{0.37} & \textbf{0.34} & \textbf{0.38} & 1025 & 169 & 3.08 & 213 & 958 & 9.30 & 3.00 & 26087\\
\hline

\end{tabular}}
\caption{Performance comparison of techniques across various metrics. Each subtable corresponds to a specific use case, with the final subtable summarizing average metric values across all case studies. Bold entries indicate the lowest values for each metric. Columns representing control-flow generation metrics are highlighted in green, time-related metrics in red, congestion-related in pink, and finally resource-related metrics in yellow. Some values of CWD are missing since the related methods only generate roles. Some values for case studies Sepsis, RTF and BPI19 are not available since the process data do not report the start timestamp, making DSIM, LSTM, LSTM(GAN), RIMS and AgentSimulator not applicable, and the evaluation of AED, RED and CED impossible since they are related to event durations.}
\label{tab:results_distances_cfecc}
\end{table}

\begin{table}[t!]
\vspace{1cm}
    \centering
    \begin{tabular}{|c||c|c|c|}
    \hline
        Use Case & \textbf{Baseline} & \textbf{SIMOD} & \textbf{CVAE} \\
        \hline\hline
        BPI17       & \textbf{1122}  & 2487   & 2373   \\
        \hline
        BPI12       & \textbf{17.5}     & 45.3     & 290   \\
        \hline
        Production  & \textbf{2.44}      & 7.91      & 8.63      \\
        \hline
        Sepsis      & 43.2     & 110    & \textbf{5.93}      \\
        \hline
        RTF         & 43.9     & 102    & \textbf{16.9}     \\
        \hline
        BPI19       & \textbf{25.8}     & 65.7     & 3502   \\
        \hline\hline
        \textbf{Average} & \textbf{209} & 470 & 1033 \\
        \hline
    \end{tabular}
    \caption{Comparison of Baseline, SIMOD, and CVAE in generating attributes, using the DAD metric. Consulta and Purchasing processes have been excluded since no attributes are present in them.}
    \label{tab:results_distances_attr}
\end{table}

As shown by the average results in Table~\ref{tab:results_distances_cfecc}, the baseline, RIMS and CVAE techniques demonstrate a general superior performance compared to all others, achieving the lowest average distances in all proposed metrics. In particular, CVAE leads in all three control-flow distances, while obtaining comparable results with the baseline approach for CFLD.
On the other hand, RIMS outperforms all the techniques in two out of three temporal distances. These strong performances can be attributed to its integration of deep learning models into the simulation at runtime, associated with the inter-case feature calculations and the inclusion of real-time queue information. On the other hand, CED is the only time-related metric that remains unaffected by this approach, making the baseline more effective than RIMS. For the same reason, RIMS yields the best results in congestion-related metrics (CTD and CAR), where accurate timestamps generation is crucial.
Finally, from the role and resource perspectives, as indicated by the RBCED and CWD distances, the baseline outperforms all other techniques, while the best results in terms of HWD are obtained by AgentSimulator, which is a resource-centric technique.

Table~\ref{tab:results_distances_attr} reports a comparison of the baseline technique, SIMOD and CVAE in generating attributes across the six processes containing them. The results highlight that the baseline and the CVAE technique outperform SIMOD in terms of DAD, with substantially lower values in all cases, with the baseline outperforming CVAE in four cases out of six. 
On average, the baseline technique achieved DAD values that were half those of SIMOD and one-fifth those of CVAE.


\begin{table}[t!]
\centering
\resizebox{\columnwidth}{!}{
\begin{tabular}{|c||c|c|c|c|c|c|c|c||c||c|}
\hline
Use Case  & \textbf{Baseline} & \textbf{DSIM} & \textbf{LSTM} & \textbf{LSTM (GAN)} & \textbf{RIMS} & \textbf{SIMOD} & \textbf{AgentSimulator} & \textbf{CVAE} & \textbf{Real} & \textbf{\#TestTraces}  \\
\hline
\hline
\textbf{BPI17}      & \cellcolor{gray!20}12.88\%  & \cellcolor{gray!20}20.58\%   & \cellcolor{gray!20}61.59\% & \cellcolor{gray!20}57.05\%     & \cellcolor{gray!20}\textbf{11.77\%}   & \cellcolor{gray!20}22.64\%  & \cellcolor{gray!20}20.68\%  & \cellcolor{gray!20} 58.33\%  & 13.45\%  & 24221  \\ \hline
\textbf{BPI12}      & \cellcolor{gray!20}\textbf{23.55\%}  & \cellcolor{gray!20}33.18\%   & \cellcolor{gray!20}80.37\% & \cellcolor{gray!20}87.97\%     & \cellcolor{gray!20}24.07\%    & \cellcolor{gray!20}25.34\% & \cellcolor{gray!20}29.78\% & \cellcolor{gray!20} 54.47\% & 20.09\%  & 6892  \\ \hline
\textbf{Production} & \cellcolor{gray!20}32.60\%  & \cellcolor{gray!20}57.20\%    & \cellcolor{gray!20}63.80\%  & \cellcolor{gray!20}72.59\%     & \cellcolor{gray!20}\textbf{32.29\%}   & \cellcolor{gray!20}33.47\% & \cellcolor{gray!20}34.57\%  & \cellcolor{gray!20} 148.38\% & 33.60\%  & 180   \\ \hline
\textbf{Purchasing} & \cellcolor{gray!20}\textbf{17.29\%}  & \cellcolor{gray!20}33.55\%   & \cellcolor{gray!20}36.24\% & \cellcolor{gray!20}81.82\%     & \cellcolor{gray!20}19.09\%   & \cellcolor{gray!20}20.72\% & \cellcolor{gray!20}21.57\%  & \cellcolor{gray!20} 80.15\% & 16.79\%  & 486     \\ \hline
\textbf{Consulta}   & \cellcolor{gray!20}\textbf{44.59\%}  & \cellcolor{gray!20}63.89\%   & \cellcolor{gray!20}62.12\% & \cellcolor{gray!20}72.61\%     & \cellcolor{gray!20}45.43\%   & \cellcolor{gray!20}47.65\% & \cellcolor{gray!20}47.26\%  & \cellcolor{gray!20} 139.89\% & 40.06\%  & 763     \\ \hline
\textbf{Sepsis}   & \cellcolor{gray!20} \textbf{28.11\%}  & \cellcolor{gray!20}-& \cellcolor{gray!20}-& \cellcolor{gray!20}-& \cellcolor{gray!20}-& \cellcolor{gray!20}62.71\% & \cellcolor{gray!20}-& \cellcolor{gray!20}87.44\% & 20.99\%  & 400     \\ \hline

\textbf{RTF}   & \cellcolor{gray!20}\textbf{42.15\%}  & \cellcolor{gray!20}-& \cellcolor{gray!20}-& \cellcolor{gray!20}-& \cellcolor{gray!20}-& \cellcolor{gray!20}59.41\% & \cellcolor{gray!20}-& \cellcolor{gray!20}40.15\% & 37.41\%  & 45,11     \\ \hline

\textbf{BPI19}   & \cellcolor{gray!20}\textbf{59.33\%} & \cellcolor{gray!20}-& \cellcolor{gray!20}-& \cellcolor{gray!20}-& \cellcolor{gray!20}-& \cellcolor{gray!20}58.15\% & \cellcolor{gray!20} - & \cellcolor{gray!20}87.59\% & 56.65\%  & 1559    \\ \hline
\hline

\hline
\textbf{Average} &
\cellcolor{gray!20}32.56\% &
\cellcolor{gray!20}41.68\% &
\cellcolor{gray!20}60.82\% &
\cellcolor{gray!20}74.41\% &
\cellcolor{gray!20}\textbf{26.53\%} &
\cellcolor{gray!20}41.26\% &
\cellcolor{gray!20}30.77\% &
\cellcolor{gray!20}87.05\% &
29.88\% &
- \\ \hline

\end{tabular}}
\caption{Relative MAE for the experiments in which the ML model has been trained on generated synthetic event data and tested on the real event data. The goal of the ML is to predict the total execution time of process instances (a.k.a.\ cycle time). In each row, the lowest values have been highlighted in bold, excluding the \textit{Real} column. The column \textit{Real} provides an easy comparison of the performance metrics when both training and testing is done on real-life data, while the column \textit{\#TestTraces} provides statistics on the number of traces of the test logs.}
\label{tab:train_on_syntetic_test}
\end{table}

\subsubsection{Predictions Quality Preservation with Syntetic Training}\label{pred}
\noindent
This section focuses on the comparison of the different techniques using the \textit{Train-on-Synthetic-Test-on-Real} method introduced in Section~\ref{subsec:generation-metrics}. Results are reported in Table~\ref{tab:train_on_syntetic_test} for the different state-of-the-art techniques across the various case studies. As defined in Section~\ref{subsec:generation-metrics}, the results reported in the table are reported in terms of Relative Mean Absolute Error, aiming to predict the total time of the trace. 
The Baseline and RIMS methods consistently outperform alternative approaches, frequently achieving performance comparable to or exceeding that obtained when the techniques are trained on real data. Although the Baseline method exhibits a higher mean accuracy than RIMS (26.18\% when evaluated on the same event logs), both methods demonstrate strong and stable results across datasets. For BPI17, RIMS achieves the highest performance (11.77\%), surpassing the baseline value that is 12.88\%, while for a similar process, BPI12, the baseline gets an rMAE of 23.55\%, however, RIMS' score is close behind (26.13\%). For Production, RIMS exhibits the best performance (32.99\%), again followed by the baseline, obtaining a value of 32.60\%.
In the Purchasing use case, the baseline marginally outperforms RIMS, achieving a value of 17.29\% and 19.09\%, respectively. Lastly, in the  Consulta dataset all the rMAE values were consistently high, the baseline has an rMAE of 44.59\%, closely followed by RIMS 45.53\%. Techniques such as DSIM, LSTM, and SIMOD demonstrate moderate performance but generally lag, with CVAE and LSTM (GAN) demonstrating their scarce ability in generating times, as highlighted in Table~\ref{tab:results_distances_cfecc}. 

This analysis highlights RIMS and the baseline as the most effective techniques across datasets for these tasks. Their superior performance is consistent with the results obtained in Table~\ref{tab:results_distances_cfecc}. In fact, these two methods showed the best results in modelling time and congestion-related features, that are the most influential factors in predicting total execution time.

\subsubsection{Log Variability through Entropy}\label{subsubsec:entropy_res}\noindent
Tables~\ref{tab:entropy_prefix} and~\ref{tab:entropy_times} report on the results of the entropy-based analysis, respectively focusing on the variability of control-flow (through Trace and Prefix entropy) and the time perspective (through Cycle Time and Activity Duration Entropy). These measures aim to assess the level of generalization in generation, aiming to generate traces that were not present in the original event log, enhancing the information contained in the original dataset, as introduced in Section~\ref{subsubsec:entropy_measures}.

The baseline technique achieves the highest values for both trace and prefix entropy across most datasets, indicating a greater diversity in the generated execution sequences. This suggests that the baseline technique effectively avoids overfitting to the most frequent patterns of the training data and is capable of generating realistic but varied traces. RIMS also performs well in this context, although slightly lower in entropy compared to the baseline.

Notably, the baseline model consistently achieves higher entropy values for both Cycle Time distribution and Activity Duration, indicating better generalization compared to other models across the logs. While the CVAE model sometimes matches or exceeds the baseline in Activity Duration Entropy, its poor performance in time and congestion distance metrics suggests that this may be due to underfitting in the temporal dimension rather than superior generalization.

\begin{table}[t]
\centering
\resizebox{\textwidth}{!}{
\begin{tabular}{|l||cc|cc|cc|cc|cc|cc|cc|cc|ll|}
\hline
\multirow{2}{*}{Use Case}  & \multicolumn{2}{c|}{\textbf{Baseline}} & \multicolumn{2}{c|}{\textbf{DSIM}} & \multicolumn{2}{c|}{\textbf{LSTM}} & \multicolumn{2}{c|}{\textbf{LSTM (GAN)}} & \multicolumn{2}{c|}{\textbf{RIMS}} & \multicolumn{2}{c|}{\textbf{SIMOD}} & \multicolumn{2}{c|}{\textbf{AgentSimulator}} & \multicolumn{2}{c|}{\textbf{CVAE}} &  \multicolumn{2}{c|}{\textbf{Real}}\\

& Trace & Prefix & Trace & Prefix & Trace & Prefix & Trace & Prefix & Trace & Prefix & Trace & Prefix & Trace & Prefix & Trace & Prefix  & Trace &Prefix \\
\hline
\hline
Sepsis      & \cellcolor{gray!20}\textbf{0.98} & \cellcolor{gray!20}\textbf{0.93} & \cellcolor{gray!20}-     & \cellcolor{gray!20}-     & \cellcolor{gray!20}-    & \cellcolor{gray!20}-    & \cellcolor{gray!20}- & \cellcolor{gray!20}- & \cellcolor{gray!20}-     & \cellcolor{gray!20}-     & \cellcolor{gray!20}0.95 & \cellcolor{gray!20}0.91 & \cellcolor{gray!20}-     & \cellcolor{gray!20}-     & \cellcolor{gray!20}0.96 & \cellcolor{gray!20}\textbf{0.93}  & 0.97 &0.89 \\
Purchasing  & \cellcolor{gray!20}\textbf{1.28} & \cellcolor{gray!20}1.23 & \cellcolor{gray!20}1.25  & \cellcolor{gray!20}1.21  & \cellcolor{gray!20}1.21 & \cellcolor{gray!20}1.14 & \cellcolor{gray!20}0.07 & \cellcolor{gray!20}0.05 & \cellcolor{gray!20}1.25  & \cellcolor{gray!20}1.19  & \cellcolor{gray!20}\textbf{1.28} & \cellcolor{gray!20}1.208 & \cellcolor{gray!20}1.27  & \cellcolor{gray!20}\textbf{1.26}  & \cellcolor{gray!20}1.25 & \cellcolor{gray!20}1.16  & \textbf{1.28} &1.20 \\
BPI17       & \cellcolor{gray!20}0.64 & \cellcolor{gray!20}0.69 & \cellcolor{gray!20}0.76  & \cellcolor{gray!20}0.76  & \cellcolor{gray!20}0.76 & \cellcolor{gray!20}0.76 & \cellcolor{gray!20}0.02 & \cellcolor{gray!20}0.03 & \cellcolor{gray!20}\textbf{0.80}  & \cellcolor{gray!20}\textbf{0.78} & \cellcolor{gray!20}0.75 & \cellcolor{gray!20}0.73 & \cellcolor{gray!20}0.63  & \cellcolor{gray!20}0.63  & \cellcolor{gray!20}0.62 & \cellcolor{gray!20}0.62  & 0.64 &0.65 \\
Production  & \cellcolor{gray!20}\textbf{1.30} & \cellcolor{gray!20}\textbf{1.25} & \cellcolor{gray!20}1.16  & \cellcolor{gray!20}1.18  & \cellcolor{gray!20}1.18 & \cellcolor{gray!20}1.16 & \cellcolor{gray!20}0.04 & \cellcolor{gray!20}0.05 & \cellcolor{gray!20}1.14  & \cellcolor{gray!20}1.15 & \cellcolor{gray!20}0.82 & \cellcolor{gray!20}0.89 & \cellcolor{gray!20}1.12  & \cellcolor{gray!20}1.11  & \cellcolor{gray!20} 0.43 &\cellcolor{gray!20} 0.29  & 1.30 &1.23 \\
BPI19       & \cellcolor{gray!20}\textbf{0.97} & \cellcolor{gray!20}\textbf{0.88} & \cellcolor{gray!20}-     & \cellcolor{gray!20}-     & \cellcolor{gray!20}-    & \cellcolor{gray!20}-    & \cellcolor{gray!20}- & \cellcolor{gray!20}- & \cellcolor{gray!20}- & \cellcolor{gray!20}-     & \cellcolor{gray!20}0.86 & \cellcolor{gray!20}0.57 & \cellcolor{gray!20}- & \cellcolor{gray!20}- & \cellcolor{gray!20}0.88 & \cellcolor{gray!20}0.86  & 0.92 &0.53 \\
Consulta    & \cellcolor{gray!20}1.07 & \cellcolor{gray!20}1.00 & \cellcolor{gray!20}0.97  & \cellcolor{gray!20}0.87  & \cellcolor{gray!20}0.97 & \cellcolor{gray!20}0.87 & \cellcolor{gray!20}0.06 & \cellcolor{gray!20}0.07 & \cellcolor{gray!20}1.03  & \cellcolor{gray!20}0.93  & \cellcolor{gray!20}0.49 & \cellcolor{gray!20}0.488 & \cellcolor{gray!20}1.07  & \cellcolor{gray!20}\textbf{1.04} & \cellcolor{gray!20}\textbf{1.08} & \cellcolor{gray!20}0.97  & 1.07 &0.99 \\
RTF         & \cellcolor{gray!20}\textbf{1.15} & \cellcolor{gray!20}\textbf{1.13} & \cellcolor{gray!20}- & \cellcolor{gray!20}- & \cellcolor{gray!20}- & \cellcolor{gray!20}- & \cellcolor{gray!20}- & \cellcolor{gray!20}- & \cellcolor{gray!20}- & \cellcolor{gray!20}- & \cellcolor{gray!20}0.92 & \cellcolor{gray!20}0.87 & \cellcolor{gray!20}- & \cellcolor{gray!20}- & \cellcolor{gray!20}0.78 & \cellcolor{gray!20}0.69  & 1.05 &1.01 \\
\hline\hline 

Average & \cellcolor{gray!20}\textbf{1.09} & \cellcolor{gray!20}\textbf{1.033} & \cellcolor{gray!20}0.975 & \cellcolor{gray!20}0.91 & \cellcolor{gray!20}0.957 & \cellcolor{gray!20}0.95 & \cellcolor{gray!20}0.04 & \cellcolor{gray!20}0.045 & \cellcolor{gray!20}0.994 & \cellcolor{gray!20}0.96 & \cellcolor{gray!20}0.764 & \cellcolor{gray!20}0.754 & \cellcolor{gray!20}0.958 & \cellcolor{gray!20}0.947 & \cellcolor{gray!20}0.882 & \cellcolor{gray!20}0.813 & 0.991 &0.894 \\
\hline

\end{tabular}}
\caption{Combined entropy values for each model across logs using both Trace and Prefix-based evaluation. Best values per use case have been highlighted in bold.}
\label{tab:entropy_prefix}
\end{table}

\begin{table}[t]
\centering
\resizebox{\linewidth}{!}{
\begin{tabular}{|l||cc|cc|cc|cc|cc|cc|cc|cc|cc|}
\hline
\multirow{2}{*}{Use Case}  
& \multicolumn{2}{c|}{\textbf{Baseline}}  
& \multicolumn{2}{c|}{\textbf{DSIM}}  
& \multicolumn{2}{c|}{\textbf{LSTM}}  
& \multicolumn{2}{c|}{\textbf{LSTM (GAN)}}  
& \multicolumn{2}{c|}{\textbf{RIMS}}  
& \multicolumn{2}{c|}{\textbf{AgentSimulator}}  
& \multicolumn{2}{c|}{\textbf{SIMOD}}  
& \multicolumn{2}{c|}{\textbf{CVAE}}  
& \multicolumn{2}{c|}{\textbf{Real}} \\
& Act & Trace & Act & Trace & Act & Trace & Act & Trace 
& Act & Trace & Act & Trace & Act & Trace & Act & Trace & Act & Trace \\
\hline\hline
BPI12       & \cellcolor{gray!20} 2.56 & \cellcolor{gray!20} \textbf{2.69} & \cellcolor{gray!20} 0.44 & \cellcolor{gray!20} 1.49 & \cellcolor{gray!20} 1.51 & \cellcolor{gray!20} 2.04 & \cellcolor{gray!20} 1.34 & \cellcolor{gray!20} 2.72 & \cellcolor{gray!20} 0.51 & \cellcolor{gray!20} 1.62 & \cellcolor{gray!20} \textbf{2.61} & \cellcolor{gray!20} 2.62 & \cellcolor{gray!20} 1.71 & \cellcolor{gray!20} 2.90 & \cellcolor{gray!20} 2.19 & \cellcolor{gray!20} 2.12 & 1.98 & 2.09 \\
Sepsis      & \cellcolor{gray!20} -    & \cellcolor{gray!20} \textbf{2.87} & \cellcolor{gray!20} -    & \cellcolor{gray!20} -    & \cellcolor{gray!20} -    & \cellcolor{gray!20} -    & \cellcolor{gray!20} -    & \cellcolor{gray!20} -    & \cellcolor{gray!20} -    & \cellcolor{gray!20} -    & \cellcolor{gray!20} -    & \cellcolor{gray!20} -    & \cellcolor{gray!20} -    & \cellcolor{gray!20} 2.12 & \cellcolor{gray!20} - & \cellcolor{gray!20} 2.02 & -    & 1.95 \\
Purchasing  & \cellcolor{gray!20} 1.76 & \cellcolor{gray!20} \textbf{2.79} & \cellcolor{gray!20} 1.18 & \cellcolor{gray!20} 1.78 & \cellcolor{gray!20} 0.64 & \cellcolor{gray!20} 1.20 & \cellcolor{gray!20} 0.98 & \cellcolor{gray!20} 1.72 & \cellcolor{gray!20} 1.04 & \cellcolor{gray!20} 1.58 & \cellcolor{gray!20} 1.65 & \cellcolor{gray!20} 1.84 & \cellcolor{gray!20} \textbf{2.74} & \cellcolor{gray!20} 2.02 & \cellcolor{gray!20} 1.65 & \cellcolor{gray!20} 2.36 & 1.33 & 1.25 \\
BPI17       & \cellcolor{gray!20} \textbf{2.47} & \cellcolor{gray!20} \textbf{3.71} & \cellcolor{gray!20} 0.59 & \cellcolor{gray!20} 2.85 & \cellcolor{gray!20} 0.69 & \cellcolor{gray!20} 3.37 & \cellcolor{gray!20} 2.20 & \cellcolor{gray!20} 3.34 & \cellcolor{gray!20} 0.27 & \cellcolor{gray!20} 3.20 & \cellcolor{gray!20} 2.20 & \cellcolor{gray!20} 2.82 & \cellcolor{gray!20} 1.95 & \cellcolor{gray!20} 3.95 & \cellcolor{gray!20} 2.25 & \cellcolor{gray!20} 3.23 & 1.82 & 3.08 \\
Production  & \cellcolor{gray!20} 1.33 & \cellcolor{gray!20} \textbf{2.24} & \cellcolor{gray!20} 1.48 & \cellcolor{gray!20} 1.79 & \cellcolor{gray!20} 0.46 & \cellcolor{gray!20} 0.00 & \cellcolor{gray!20} 0.70 & \cellcolor{gray!20} 1.70 & \cellcolor{gray!20} 1.53 & \cellcolor{gray!20} 1.46 & \cellcolor{gray!20} 1.32 & \cellcolor{gray!20} 1.05 & \cellcolor{gray!20} 0.56 & \cellcolor{gray!20} 0.96 & \cellcolor{gray!20} \textbf{1.83} & \cellcolor{gray!20} 0.98 & 1.00 & 1.74 \\
BPI19       & \cellcolor{gray!20} -    & \cellcolor{gray!20} \textbf{4.03} & \cellcolor{gray!20} -    & \cellcolor{gray!20} -    & \cellcolor{gray!20} -    & \cellcolor{gray!20} -    & \cellcolor{gray!20} -    & \cellcolor{gray!20} -    & \cellcolor{gray!20} -    & \cellcolor{gray!20} -    & \cellcolor{gray!20} -    & \cellcolor{gray!20} -    & \cellcolor{gray!20} -    & \cellcolor{gray!20} 1.87 & \cellcolor{gray!20} - & \cellcolor{gray!20} 0.89 & -    & 2.05 \\
Consulta    & \cellcolor{gray!20} 1.14 & \cellcolor{gray!20} \textbf{2.67} & \cellcolor{gray!20} 0.71 & \cellcolor{gray!20} 1.75 & \cellcolor{gray!20} 0.00 & \cellcolor{gray!20} 1.07 & \cellcolor{gray!20} 0.65 & \cellcolor{gray!20} 1.86 & \cellcolor{gray!20} 0.63 & \cellcolor{gray!20} 1.78 & \cellcolor{gray!20} 1.25 & \cellcolor{gray!20} 1.28 & \cellcolor{gray!20} 0.61 & \cellcolor{gray!20} 1.08 & \cellcolor{gray!20} \textbf{1.29} & \cellcolor{gray!20} 1.91 & 0.73 & 1.71 \\
RTF         & \cellcolor{gray!20} -    & \cellcolor{gray!20} \textbf{2.90} & \cellcolor{gray!20} -    & \cellcolor{gray!20} -    & \cellcolor{gray!20} -    & \cellcolor{gray!20} -    & \cellcolor{gray!20} -    & \cellcolor{gray!20} -    & \cellcolor{gray!20} -    & \cellcolor{gray!20} -    & \cellcolor{gray!20} -    & \cellcolor{gray!20} -    & \cellcolor{gray!20} -    & \cellcolor{gray!20} 2.01 & \cellcolor{gray!20} -    & \cellcolor{gray!20} 0.98 & -    & 2.53 \\
\hline\hline
\textbf{Average} 
& \cellcolor{gray!20} \textbf{1.85} & \cellcolor{gray!20} \textbf{2.99} & \cellcolor{gray!20} 0.88 & \cellcolor{gray!20} 1.93 & \cellcolor{gray!20} 0.66 & \cellcolor{gray!20} 1.54 & \cellcolor{gray!20} 1.17 & \cellcolor{gray!20} 2.27 & \cellcolor{gray!20} 0.80 & \cellcolor{gray!20} 1.93 & \cellcolor{gray!20} 1.81 & \cellcolor{gray!20} 1.92 & \cellcolor{gray!20} 1.51 & \cellcolor{gray!20} 2.11 & \cellcolor{gray!20} 1.77 & \cellcolor{gray!20} 1.81 & 1.37 & 2.05 \\
\hline
\end{tabular}
}
\caption{Combined entropy values for each model across logs using both Cycle Time and Activity Duration entropies. Values for Activity Duration Entropy have not been reported for logs that do not contain start timestamps. Best values per use case have been highlighted in bold.}
\label{tab:entropy_times}
\end{table}

\subsubsection{Comparison of Computational Times}\label{times}
\begin{table}[t!]
\centering
\resizebox{\columnwidth}{!}{
\begin{tabular}{|l|c|c|c|c|c|c|c|c|}
\hline
Use Case        & \textbf{Baseline} & \textbf{DSIM} & \textbf{LSTM} & \textbf{LSTM(GAN)} & \textbf{RIMS} & \textbf{SIMOD} & \textbf{AgentSimulator} & \textbf{CVAE} \\ \hline
\hline
\textbf{BPI17}       &  1min 30s & 2h 30min & 9h 30min  & 10h 45min & >DSIM & 1h 5min  & 50min    & 14h 56min   \\ \hline
\textbf{BPI12}       &  14s      & 2h 0min  & 7h 30min  & 7h 40min  & >DSIM & 55min    & 22min    & 3h 31min    \\ \hline
\textbf{Production}  &  16s      & 55min    & 2h 30min  & 2h 45min  & >DSIM & 48min    & 55s      & 30min 00s   \\ \hline
\textbf{Purchasing}  &  5s       & 1h 10min & 2h 45min  & 3h 30min  & >DSIM & 45min    & 1min 50s & 34min 18s   \\ \hline
\textbf{Consulta}    &  14s      & 50min    & 3h 45min  & 4h 20min  & >DSIM & 42min    & 1min 50s & 35min 38s   \\ \hline
\textbf{BPI19}       &  45s      & -        & -         & -         & -     & 1h 34min & 1h 2min  & 8h 33min    \\ \hline
\textbf{RTF}         &  1min 50s & -        & -         & -         & -     & 3h 45min & 1h 40min & 17h 33min   \\ \hline
\textbf{SEPSIS}      &  12s      & -        & -         & -         & -     & 45min    & 1h 20min & 59min 35s   \\ \hline
\hline
\textbf{Average} &  29s & 1h 34min & 5h 12min & 5h 48min & >DSIM & 1h 17min & 41min & 6h 17min \\
\hline
\end{tabular}}
\caption{Comparison of time performance across different case studies and methods. Times for RIMS are not reported, as its training procedure relies on DSIM training.}
\label{tab:times}
\end{table}

Table~\ref{tab:times} displays the training and generation times for each use case. The table shows that the baseline technique demonstrates the lowest computational times across all the case studies, with times ranging from 0 to 3 minutes. This suggests its efficiency in handling every scenario due to the fact that it is solely based on probability sampling. In contrast, DSIM, LSTM exhibit significantly longer processing times, particularly for complex processes such as BPI17, which present more traces and complexity. RIMS time have not been reported, due to its dependency of DSIM training (cf.\ Section~\ref{sec:related_works}) that makes its computational times higher than them.
The introduction of GANs to LSTM further increases the computational load, with LSTM (GAN) showing the longest times among all methods, reaching up to 12 hours. On the other hand, AgentSimulator is able to reach low computational time for small logs while keeping moderately low computational values also for bigger log as RTF and BPI19. Similarly, SIMOD shows moderate computational times but performs notably faster than DSIM, RIMS, LSTM and LSTM (GAN), being also able to generate resources and attributes.

As reported in Section~\ref{subsec:use_cases}, computational times for BPI19, RTF and SEPSIS have only been computed for the baseline, AgentSimulator, SIMOD and CVAE, since they lack complete timestamps, that are necessary for the other techniques.

These results underscore the trade-off between computational efficiency and the complexity of the underlying techniques. While advanced techniques such as GANs can provide sophisticated insights, their applicability might be limited by the extensive computational resources and time required, especially if applied to process mining encodings, which usually require a limited number of features to encode. \emph{The baseline and AgentSimulator, with their relatively low computational times, present themselves as viable options for real-time or near-real-time process mining applications.}

Furthermore, these findings highlight the scalability of the baseline and AgentSimulator, which maintain stable results even varying log sizes while requiring minimal computational resources. 

\subsubsection{Synthesis and Interpretation of the Findings}\noindent
\label{sec:discussion}
Table~\ref{tab:model_comparison} summarizes the findings related to this Section. From the second to the sixth column of the table the results related to the \textit{similarity} are reported, in particular related to the perspective of control-flow (CF), time, congestion (Cong), resources (Res) and attributes (Attr), while the following columns reports results pertaining to the \textit{data quality preservation}, measured through the Train-on-Syntetic-Test-on-Real (TSTR), the \textit{log variability} measured through entropy and the \textit{computational times}.
For each column, the star symbol \ding{72} indicates the technique that scores the best for a corresponding criterion, while a percentage value indicates how much a certain criterion scores worse than the best technique, namely $(100\% \cdot (v-b)) \div b$ with $v$ being the value of the technique in question and $b$ being the value for the best technique. 
For the perspectives related to the similarity, the reported results represent the average of the individual percentual improvement observed for each metric in Table~\ref{tab:results_distances_cfecc}. This technique was adopted due to the differing scales of the reported quantities. Specifically, for each metric within a given perspective, the percentual improvement $(100\% \cdot (v-b)) \div b$ achieved by each model with respect to the best one in that perspective was first computed. This procedure ensures that the comparison reflects relative performance gains rather than absolute differences, thereby eliminating the influence of heterogeneous scales or measurement units across metrics. Subsequently, these percentual improvements were aggregated by averaging them within each perspective, yielding a single composite indicator representing the overall improvement in a scale-independent manner. This aggregation strategy enables a fair and methodologically sound comparison between models, as it mitigates potential distortions arising from disparities in metric magnitudes or units of measurement.

\begin{table}[t!]
\centering
\vspace{1cm}
        \resizebox{\columnwidth}{!}{
            \begin{tabular}{lcccccccc}
                \toprule
                \textbf{Technique}      &  \textbf{CF} &\textbf{Time}& \textbf{Cong} & \textbf{Res} & \textbf{Attr} &\textbf{TSTR} & \textbf{Log Variability} & \textbf{Computation Time} \\
                \midrule
                \textbf{Baseline}       & 16\%         &1581\%     &104\%       & 94\%      & \ding{72} & 23\%     & \ding{72}   & \ding{72} \\ 
                \textbf{DSIM}           & 74\%         &560\%      &595\%       & -         & -         & 57\%     & -26\%       & 19348\%    \\ 
                \textbf{LSTM}           & 87\%         &957\%      &531\%       & -         & -         & 130\%    & -32\%       & 64451\%   \\ 
                \textbf{LSTM (GAN)}     & 145\%        &33175\%    &84135\%     & -         & -         & 184\%    & -62\%       & 71900\%   \\ 
                \textbf{RIMS}           & 78\%         &\ding{72}  &\ding{72}   & -         & -         & \ding{72}& -26\%       & >71900\%   \\ 
                \textbf{SIMOD}          & 54\%         &18689\%    &12349\%     & 170\%     & 124\%     & 57\%     & -25\%       & 15831\%   \\ 
                \textbf{AgentSimulator} & 69\%         &292\%      &499\%       & \ding{72} & -         & 15\%     & -17\%       & 8382\%    \\ 
                \textbf{CVAE}           & \ding{72}    &387\%      &429\%       & 37\%      & 394\%     & 234\%    & -23\%       & 77900\%   \\ 
                \bottomrule
            \end{tabular}}
    \caption{Comparison of techniques across various metrics. For each column, the cell with \ding{72} is the technique that scores the best, while the percentage values indicate the extent with which each technique is worse than the best. Column \textit{CF, Time, Cong, Res, Attr} refer respectively to the control-flow, time, congestion, resources and attributes perspectives reported in Tables~\ref{tab:results_distances_cfecc} and~\ref{tab:results_distances_attr}, \textit{TSTR} refers to the metrics of \textit{Train-on-Synthetic-Test-on-Real}.}
\label{tab:model_comparison}
\end{table}

The metrics related to the control-flow similarity of the original and generated event log shows that CVAE is able to better mimic the original-log benavior into the generated event log, with the baseline being 16\% better. This superiority of the CVAE was expected, as already observed in~\cite{DBLP:conf/icpm/GraziosiRBFFGMP24,Graziosi2025} and in other domains, such as music generation (see, e.g.,~\cite{Comanducci2023}). Nevertheless, the baseline also performs reasonably well. This may be because event sequences (i.e., traces) often have a simple structure, where the most recent few events effectively summarize the prior history, thus providing a strong state abstraction.

RIMS achieves the best performance with respect to time- and congestion-related metrics. This result is unsurprising, as RIMS explicitly models complex queuing systems, which naturally excel at capturing the factors influencing execution and waiting times of activities. In terms of resource-related metrics (such as HWD), AgentSimulator performs significantly better than CVAE and the baseline, which are the only other techniques, apart from AgentSimulator, that model the resource perspective. This is expected, given that AgentSimulator considers the resource perspective as the primary dimension, thus prioritizing resource interactions and handovers, while all other approaches treat the control-flow perspective as their structural backbone.
From the attribute perspective, the baseline achieves the best results among the methods capable of generating attributes, whereas CVAE and SIMOD yield lower performance values. This behavior is associated with the baseline model’s ability to accurately learn the probability distribution of the global attributes and subsequently assign them to a trace after its generation, which results in an EMD's value closer to zero. Conversely, SIMOD operates as a simulator-based generator that relies on deterministic and stochastic attribute update rules, as well as data-aware branching conditions. While this continuous updating mechanism enhances the realism of the simulation process, it tends to reduce the model’s accuracy in estimating an explicit probability distribution. Moreover, the CVAE employs an autoencoder-based architecture that first estimates a latent distribution from which attribute values are sampled, adding one further step to the probability estimation process that affects the value of EMD. It is noteworthy, however, that in scenarios characterized by overfitting or concept drift, the CVAE may outperform other approaches due to its capacity to adapt its latent space representation to evolving data patterns.

The column TSTR highlights that the Train-on-Synthetic-Test-on-Real  metrics shows that RIMS generates synthetic data with the highest utility to train a Machine- or Deep-Learning model, followed by Agent Simulator (15\% lower value) and the baseline (23 \% lower). This is certainly related to the previous observation that RIMS is the best to to generate the time- and congestion-related metrics, which are highlightly related to the prediction task for which training and test were carried on: predicting the process-instance cycle time. Since cycle time is also influenced by resource availability, this explains why AgentSimulator performs well on this metric, as it excels in modeling resource dynamics. The baseline also achieves good results, likely due to its capacity to ensure higher variability (i.e., more generalizability) in the generated event logs while maintaining consistency in the control-flow perspective generation. while maintaining strong control-flow consistency. Surprisingly, CVAE performs poorly in the TSTR experiment, suggesting that CVAE produces the event logs that are not rich or realistic enough for models to generalize effectively to real data. This limitation is consistent with CVAE's weaker performance in time- and congestion-related metrics.
It should be noted that the Train-on-Synthetic-Test-on-Real  metric was evaluated exclusively, though across several processes and logs, in the context of prediction tasks related to process-instance cycle time. Different conclusions might be potentially drawn if the prediction task concerned other performance indicators or process outcomes~\cite{DiFrancescomarino2022}. Future work can go along the direction of experimenting this. 

When considering the quality of log variability, however, the baseline demonstrates a greater capacity to generalize, producing new traces not observed in the original data. It outperforms all other methods, including CVAE, likely because the number $k$ of past events used to represent the current state can be easily optimized (cf.~\ref{appendix:probagen_k}) using a validation-based optimization approach. The remaining methods exhibit similar levels of generalization, with entropy values approximately 20–30\% lower than the baseline, except for LSTM (GAN), which performs markedly worse.

However, while RIMS shows significant benefits, it is characterized by long training times (multiple hours), which is a drawback that is shared with other techniques based on simulation and/or deep-learning models, including DSIM, LSTM, GAN. Conversely, the baseline always took at most two minutes in all case studies, where AgentSimulator often took less than two minutes, with the notable exceptions of the BPI17 and BPI12 case studies that led to a significant increase of the average value.

In conclusion, the experiments illustrate that the problem of event-log generation can be tackled in practice, and that a simple baseline is already capable to provide good outcomes. More advanced techniques are capable to generate even better synthetic event-logs, which better capture certain process perspectives. For example, RIMS excels in generating the time-related aspects of the event logs because of its detailed queue-based modelling of the resource perspective. 
An interesting direction of future work could be to integrate RIMS and the baseline. The baseline guarantees an accurate and fast generation of the event-log control-flow sequences that is also provided to be very useful to train prediction models, while RIMS excels at discovering the resource queues that affect the time perspective. An integrated approach can likely guarantee high fidelity in event-log generation.

\section{Event-Log Augmentation for Class Balancing}\label{sec:aug_for_accuracy}\noindent
This section investigates the use of event-log augmentation to mitigate class imbalance, a common issue that hampers the prediction of infrequent activities in process models. Section~\ref{subsec:class_pred_metric} introduces the methodology used, while Section~\ref{subsec:aug_res} reports on the results.

\subsection{Methodology}\label{subsec:class_pred_metric}\noindent
The goal of this analysis is to illustrate a practical scenario in which \emph{event-log augmentation} proves to be beneficial:  balancing class distributions to improve the prediction performance for infrequent activities. The evaluation procedure follows the testing approach proposed by~\cite{8953317}. For each log, the least frequent activity $a$ is identified and extracted. Then, the framework proposed by~\cite{bib:padella} is trained on $\ML^{train}$ to predict whether, given a running trace from the same dataset used for TSTR ($\ML^{run}$), the activity $a$ will occur in the remaining control-flow, thus formulating the task as a binary classification problem. 

Among the compared methods, those achieving the highest accuracy in at least one of the perspectives reported in Table~\ref{tab:model_comparison}: the baseline technique, RIMS, CVAE, and AgentSimulator.
These four techniques were employed to generate synthetic traces containing activity $a$. The generated traces were then incorporated into $\ML^{train}$, replacing an equivalent number of traces that did not include $a$, thereby producing the augmented dataset $\ML^{train\_aug}$. In this augmented dataset, additional traces were added to ensure that activity $a$ appears in 50\% of the traces.
The same predictive framework from~\cite{bib:padella} is subsequently retrained on $\ML^{train\_aug}$ and evaluated on $\ML^{run}$. Since the objective is to assess the model’s ability to predict infrequent activities, performance is measured in terms of the \emph{F-Score}, reflecting the balance between precision and recall for both classes. 
Additionally, the analysis includes a comparison with the \textsc{SMOTE} technique applied to the encoded event log (cf.\ Section~\ref{sec:introduction}), serving as a benchmark against a traditional data augmentation method that is not process-aware.

\subsection{Results}\label{subsec:aug_res}
\noindent
Section~\ref{sec:introduction} has indicated that event-log augmentation is beneficial in predictive process monitoring when the distribution of Key Performance Indicator (KPI) values is unbalanced towards certain values. This section reports on the experiments that confirm the expectations. Table~\ref{tab:case_study_activities} reports on the F-Score of the tasks to predict whether or not certain activities will eventually be executed. In particular, the same five case studies were employed, and, each, two infrequent activities were used to predict their occurrence. \footnote{As a matter of fact, for the Production and Consulta use case, the two most infrequent activities  were skipped because they were present only once in the whole test log. Conversely, the paper focuses on the third and the fourth least present activities.}
\begin{table}[t!]
\centering
\resizebox{\columnwidth}{!}{
\begin{tabular}{llccccccc}
\toprule
\textbf{Use Case} & \textbf{Activity} & \textbf{Freq (\%)} & \textbf{Not Balanced} & \textbf{Baseline} & \textbf{RIMS} & \textbf{CVAE} & \textbf{AgentSimulator} & \textbf{SMOTE} \\
\midrule
\textbf{BPI17} & W\_Handle leads & 11.03 & 0.55 & \textbf{0.79} & 0.70 & 0.50 & 0.72 & 0.39 \\
 & W\_Assess potential fraud & 0.74 & 0.59 & \textbf{0.71} & 0.58 & 0.50 & 0.67 & 0.32 \\
\midrule
\textbf{BPI12} & W\_Valideren aanvraag & 34.07 & 0.70 & \textbf{0.87} & 0.77 & 0.58 & 0.59 & 0.85 \\
 & W\_Beoordelen fraude & 1.23 & \textbf{0.75} & 0.75 & 0.71 & 0.58 & 0.67 & 0.60 \\
\midrule
\textbf{Production} & Flat Grinding & 26.22 & 0.45 & \textbf{0.87} & 0.77 & 0.53 & 0.80 & 0.49 \\
 & Turning & 10.67 & \textbf{0.85} & 0.83 & 0.77 & 0.49 & 0.72 & 0.80 \\
\midrule
\textbf{Purchasing} & Amend Request for Quotation & 39.14 & 0.69 & 0.64 & 0.52 & 0.55 & \textbf{0.67} & 0.24 \\
 & Settle Dispute With Supplier & 16.94 & 0.50 & \textbf{0.72} & 0.64 & 0.25 & 0.59 & 0.39 \\
\midrule
\textbf{Consulta} & Transferir créditos homologables & 5.45 & 0.47 & \textbf{0.90} & 0.84 & 0.56 & 0.68 & 0.55 \\
 & Validar solicitud pre-homologación & 7.86 & 0.40 & \textbf{0.90} & 0.87 & 0.66 & 0.74 & 0.74 \\
\midrule
\multicolumn{3}{c}{\textbf{Average}} & 0.59 & \textbf{0.79} & 0.72 & 0.52 & 0.68 & 0.56 \\
\bottomrule
\end{tabular}}
\caption{Comparison of activity prediction quality using F-Scores, with and without rebalancing the trace counts for each activity through Baseline, RIMS, CVAE, and AgentSimulator. The last column (\textit{SMOTE}) reports results for a traditional data augmentation technique applied to the encoded event log. The best score per row is highlighted in bold.}
\label{tab:case_study_activities}
\end{table}

The third column of Table~\ref{tab:case_study_activities} reports on the percentage of traces in which the activities occurred. Columns with header F-Score and F-Score balanced illustrate the values of this metric when, respectively, the original dataset was employed, or a balanced dataset was created by augmenting the infrequent class values to reach the 50\% of trace with that value. The augmentation results was made using the baseline technique, the RIMS, the CVAE and the AgentSimulator, since they proven to be the most effective frameworks on at least one perspective.

Specifically, when rebalancing the event log using a CVAE, we initially intended to exploit its conditional generation capability by labeling traces based on the presence of the target activity (cf. Section 4 in~\cite{DBLP:conf/icpm/GraziosiRBFFGMP24,Graziosi2025}). However, since the target activities were among the least frequent in the event log, the model struggled to generate a sufficient variety of traces containing them. As a result, we proceeded with an unconditional variant of the model.

Last column represents F-Score values related to the data augmentation performed applying a traditional method, i.e. SMOTE (cf.\ Section~\ref{subsec:generation-metrics}).

The findings highlight that balancing techniques based on process-aware generation, such as RIMS and the baseline probabilistic model, outperform traditional methods like SMOTE, particularly when the activity frequency is very low. SMOTE, despite being widely used in general-purpose data augmentation, struggles in capturing temporal and control-flow dependencies, which are critical in process-oriented data where events are not independent among themselves.

Interestingly, while CVAE and AgentSimulator occasionally achieved comparable performance, on the other hand, RIMS and baseline demonstrated robust improvements over the unbalanced setting, with F-Score gains of up to 30 percentage points in some cases, suggesting that incorporating simulation-based mechanisms with predictive components can be particularly effective in capturing complex inter-case features.

\section{Conclusion}
\label{sec:conclusion}
\noindent Process mining aims to analyze and improve processes by examining transactional data, which captures how individual process executions unfold. This data is structured in event logs, consisting of traces that record sequences of events, marking the start or completion of activities along with their timestamps. However, the applicability of process mining depends on the availability of sufficiently large event-logs, particularly when process mining is used in combination with techniques based on machine- or deep-learning, which require a vast quantity of training data. The need for event-log augmentation arises to overcome this limitation by generating additional traces that simulate new valid executions of a process. This augmentation process must account for multiple perspectives, including time, control-flow, resource allocation, and domain-specific attributes, ensuring that generated traces reflect realistic process behaviors.

Unlike traditional data augmentation techniques that assume stochastic independence, event-log augmentation must respect the sequential dependencies and constraints inherent in the process. Events within traces are interdependent, and resource sharing introduces dependencies across traces. Additionally, augmentation can be leveraged to generate traces of rare executions, which is valuable for predictive process monitoring in identifying infrequent but critical process behaviors. While prior studies have introduced event-log augmentation techniques, their effectiveness varies, and there is no extensive study that aims to systematically compare them. 

This paper evaluates seven state-of-the-art techniques across eight event logs, and compare them with a baseline that is based on annotated stochastic transition system. 
The experimental results show that the CVAE best generates an event log that mimics the behavior of the original logs, although the computation time makes it hard to be employed in middle-to-large case studies. The baseline also performs well due to the simplicity of event sequences. In contrast, the baseline demonstrates superior generalization in terms of log variability, while RIMS achieves the highest accuracy for time- and congestion-related metrics thanks to its explicit modeling of queuing dynamics. AgentSimulator performs best on resource-related measures, as it focuses primarily on resource interactions and handovers, and the baseline again leads among methods capable of generating attributes. Regarding utility, the Train-on-Synthetic-Test-on-Real metric indicates that RIMS produces synthetic data with the highest predictive value, followed by AgentSimulator and the baseline, while CVAE performs poorly, confirming its limited realism and weak modeling of temporal dynamics.

TSTR illustrates the utility of generating synthetic event logs, which is a different quality that fidelity. The latter can be assessed on illustrating the benefits of augmenting the event log to, e.g., tackle prediction tasks. Section~\ref{sec:aug_for_accuracy} has reported on a number of experiments where the goal was to predict whether or not certain infrequent activities were eventually going to occur during a process execution. Since the chosen activities was infrequent, the datasets were unbalanced: the four best performing techniques - RIMS, the baseline, AgentSimulator and CVAE - were used to generate additional traces so as to balance the dataset, and also compared with the results obtained via SMOTE, which a more traditional, process-unaware data augmentation method. The baseline and RIMS outperformed AgentSimulator and CVAE in terms of F-score of the prediction models trained on balanced datasets, and showed the benefits if compared with the models training on the unbalanced datasets. SMOTE did not provide any benefits, if not even reducing the F-score compared with the dataset: this has confirmed that a process-unaware augmentation method is not applicable in Process Mining. 

In summary, the study confirms that event-log augmentation is a feasible task: simple baselines already yield good results, while advanced models improve specific perspectives. Future work may focus on integrating RIMS and the baseline to combine fast and accurate control-flow generation with detailed time and resource modeling for synthetic event logs that are higher in fidelity and utility.

\section*{Declaration of generative AI and AI-assisted technologies in the manuscript preparation process}
\noindent
During the preparation of this work the authors used Grammarly and ChatGPT to assist with language refinement. After using these tools, the authors reviewed and edited the content as needed and take full responsibility for the content of the published article.

\section*{Acknowledgments}\noindent
We acknowledge the support of the project ``Future AI Research (FAIR) - Spoke 2 Integrative AI - Symbolic conditioning of Graph Generative Models (SymboliG)”  funded by the European Union under the National Recovery and Resilience Plan (NRRP), Mission 4 Component 2 Investment 1.3 - Call for tender No. 341 of March 15, 2022 of Italian Ministry of University and Research – NextGenerationEU, Code PE0000013, Concession Decree No. 1555 of October 11, 2022 CUP C63C22000770006.

\bibliographystyle{elsarticle-num}

\newpage
\appendix
\section{Probabilistic Event-Log Augmentation: Formalization}\label{appendix:probagen}
\noindent In this appendix, we present the formalization of the baseline technique described in Section~\ref{sec:framework}. The technique models event logs using \textbf{Probabilistic Transition System} that incorporates multiple process perspectives: control-flow, resources, event attributes, and temporal aspects. The goal is to estimate probability functions that describe, for a given process state, the probability of transitioning to another state.

The starting point is an event log, from which a starting transition system that captures the control-flow of the process is discovered. This is then enhanced with probabilistic functions that model the probabilities of resources and event attributes with each transition. Temporal components, such as inter-event times and inter-arrival times, are modeled using probability density functions derived from the event-log.

Given an event log $\ML$, we define a \textbf{Probabilistic Transition System} as a tuple 
\[
PTS_\ML =((S^k, T, P), (S^k_{\MR}, T_\MR, P_\MR), (S^k_{\MD}, T_\MD, P_\MD), P_\MT,s_0, S_F)
\]
with $k>0$, where each triple captures distinct process perspectives -respectively, \textit{control-flow}, \textit{resource}, and \textit{event attributes} - ,while $P_\MT$ describes the temporal perspective. The components are defined and discovered as follows:
\begin{description}
    \item[Control-Flow Discovery] $(S^k, T, P)$ describes the control-flow transition systems. Specifically, $S^k$ is the multiset of all possible sequence of $k-$prefixes of activity-lifecycle observed in the event log: $$S^k=\left[\langle(act(e),life(e))\mid e\in\sigma_k\rangle\mid\forall\sigma_k\in prefix_k(\sigma),\ \forall\sigma\in\ML\right]$$ $T\subset S^k\times(\MA\times\MI)\times S^k$ defines the set of possible transitions from a state in $S^k$ to another state in $S^k$  after performing a couple of activity-lifecycle in $\MA\times\MI$. Specifically, given $(s',(a,l),s'')\in T$, then if $s'=\langle(a_i,l_i),\dots,(a_{i+k}, l_{i+k})\rangle$ then $s''=\langle(a_{i+1},l_{i+1}),\dots,(a_{i+k}, l_{i+k}),(a,l)\rangle$. Finally, $P\colon T\to[0,1]$ describes the probability of transitioning from a state to another. Specifically, the empirical probability of observing $s''$, and hence performing $(a,l)$, after prefix $s'$ is computed as:\footnote{Symbol $\#$ indicates the frequency of elements in a multiset.}
    $$P((s',(a,l),s''))=\frac{\#[s\in S^k\mid \exists(s',(a,l),s)\in T,\ s=s'']}{\#[s\in S^k\mid\exists (s',(a,l),s)\in T]}$$
    Note that $\sum_{s}P((s',(a,l),s))=1$. 
    This control-flow part permits the generation of sequences of activity-lifecycle pairs, thus modeling the control-flow part of the process.
    \item[Resource Discovery] The resource perspective, $(S^k_{\MR}, T_\MR, P_\MR)$ extends the control-flow view by incorporating resource information. The multiset $S_\MR^k$ contains all observed sequences of $k-$prefixes including, for each event its associated resource:
    $$S_\MR^k=\left[\langle(act(e),life(e), res(e))\mid e\in\sigma_k\rangle\mid\forall\sigma_k\in prefix_k(\sigma),\ \forall\sigma\in\ML\right]$$
    The set of transitions $T_\MR\subset S^k_\MR\times(\MA\times\MI)\times S^k_\MR$ defines possible transitions between resource-aware states. Given $(s',(a,l),s'')\in T_\MR$, if $s'=\langle(a_i,l_i,r_i),\dots, \\ (a_{i+k}, l_{i+k},r_{i+k})\rangle$ then $s''=\langle(a_{i+1},l_{i+1},r_{i+1}),\dots,(a_{i+k}, l_{i+k},r_{i+k}),(a,l,r)\rangle$, with $ r\in\MR$. The probability function $P_\MR\colon T_\MR\to[0,1]$ assigns empirical probability of observing the next resource given the current prefix:
    $$P((s',(a,l), s'')) = \frac{\#[s\in S^k_\MR\mid \exists(s',(a,l),s)\in T_\MR,\ s=s'']}{\#[s\in S^k_\MR\mid\exists (s',(a,l),s)\in T_\MR]}$$
    This probability expresses the likelihood that, given a resource-aware prefix $s'\in S_\MR^k$, the next activity-lifecycle pair $(a,l)$, obtained sampling from the control-flow transition system, will be executed by a particular resource, leading to state $s''$. Note that for each $s'$ and $(a,l)$, $\sum_{s}P_\MR((s', (a,l), s))=1$.
    \item[Event Attributes Discovery] Similarly as with the resource perspective, the event attribute perspective transition systems $(S^k_{\MD}, T_\MD, P_\MD)$ extend the control-flow one , thus modeling the data dimension of the process. The multiset of states is defined as 
    $$S^k_{\MD}=\left[\langle(act(e),life(e), attr(e))\mid e\in\sigma_k\rangle\mid\forall\sigma_k\in prefix_k(\sigma),\ \forall\sigma\in\ML\right]$$
    extending the control-flow states $S^k$ by incorporating the event attributes of events. The event attributes transition set $T_\MD\subset S^k_\MD\times(\MA\times\MI)\times S^k_\MD$ links data-aware prefixes, such that $(s',(a,l), s'')\in T_\MD$ iff $s''$ results from appending $(a,l,v)$ to $s'$, where $v\in\MD$ denotes the observed attributes values. The probability function $P_\MV\colon T_\MD\to[0,1]$ assigns each transition its empirical probability:
    $$P((s',(a,l), s'')) = \frac{\#[s\in S^k_\MD\mid \exists(s',(a,l),s)\in T_\MD,\ s=s'']}{\#[s\in S^k_\MD\mid\exists (s',(a,l),s)\in T_\MD]}$$
    This probability quantifies the likelihood that, given the current attribute-aware prefix $s'$, the next activity-lifecycle $(a,l)$ will occur with specific event attributes $v$, resulting in the extended state $s''$, with $(a,l)$ the last activity-lifecycle and those attributes. Note that, for all $s'$ and $(a,l)$, $\sum_{s}P_\MD((s', (a,l), s))=1$.
    \item[Temporal Discovery] The temporal perspective models the duration associated with each activity-lifecylce pair. Formally, 
    $$P_\MT:(\MA\times\MI)\to\{d_{\Theta}\mid d\in \{C,\MN,Exp,\mathcal{U},Triang,Log\MN,\Gamma\}\}$$where each $(a,l)\in\MA\times\MI$ is mapped to a parametric distribution $d_\Theta$ from a set of predetermined ones (respectively,  \textit{constant, normal, exponential, uniform, triangular, lognormal, gamma}) with specific parameters $\Theta$ estimated from observed durations in the log. For each $(a,l)$, the corresponding distribution models the time taken to transition to state $s\in S^k$ s.t. $s=\langle\dots,(a,l)\rangle\in S^k$. Specifically, this distribution is found by collecting all the values $\{time(e_i)-time(e_{i-1})\mid \forall \sigma=\langle e_1,\dots e_n\rangle\in\ML\ s.t.\ act(e_i)=a,\ life(e_i)=l \}$ and selecting the best fitting distribution. Selection has been based on the Wasserstein distance as~\cite{villani2009wasserstein}.
\end{description}
\noindent Finally, $s_o=\langle\rangle\in S^k$ denotes the initial (empty) state, and $S_F\subseteq S^k$ is the set of final states observed in $\ML$.

The Probabilistic Transition System $PTS_\ML$ models the behavior of individual traces by capturing the likelihood of transitions. To simulate the generation of complete event logs, we also define a distribution function to model the arrival times of traces. This allows us to generate arrival of new traces over time, which are then generated using the $PTS_\ML$. The inter-arrival time distribution function $d_{AT}$ is defined similarly as the distribution functions defined by $P_\MT$: the inter-arrival times, i.e. the time between the arrival of two consecutive cases, are collected and the best fitting distribution function from a set of predetermined ones (the same as used in $P_\MT$) is selected.

Once all these components are defined, they can be used to generate event-logs. First, arrival times are generated using the inter-arrival time distribution function $d_{AT}$, starting from an initial given start timestamp, for a specified number of traces to generate. For each arrival, a corresponding trace is then generated by sampling transitions from the Probabilistic Transition System $PTL_\MS$, thus obtaining a complete event-log.

Last but not least, the parameter $k$, which determines the size of the event history considered in each state, plays a crucial role in shaping the behavior of the Probabilistic Transition System. By conditioning transitions on the last $k$ events allows not only a vanilla sampling of existing behavior but also the generation of new, feasible traces that are consistent with the underlying process dynamics.

\section{History Length Parameter Optimization}\label{appendix:probagen_k}
\noindent
Section~\ref{sec:framework} and \ref{appendix:probagen} have shown how the baseline model is based on probabilistic transition system with a hyper-parameter $k$, which indicates the number of events, counting from the last, that are considered when the state is constructed. 
The optimization of this parameter has to be analyzed under the lens of two perspectives: \textbf{similarity} and \textbf{log variability} (cf.\ Section~\ref{sec:experiments}). They are often opposing forces: higher values of $k$ lead to lower log variability and higher similarity, and vice versa. 
Therefore, we aim to choose the best $k$ to balance these two forces. Since the hyperparameter $k$ primarily influences control-flow generation, we employed the \textit{CFLD} distance (cf.\ Section~\ref{sec:metrics}) to measure similarity, and the \textit{trace entropy} to quantify log variability (cf.\ Definition~\ref{def:trace_entropy}). To ensure comparability with \textit{CFLD}, the trace entropy values were rescaled to the interval $[0,1]$ by dividing them by $\log(|\mathcal{L}|)$, that is the maximum value that it can assume. This normalization procedure follows the guidelines for entropy scaling introduced by Wilcox~\cite{Wilcox1967}.
Moreover, high values of similarity correspond to low values of CFLD, while high values of trace entropy correspond to high log variability. To treat the hyper-parameter optimization problem as a minimization problem, we considered the complementary measure of normalized trace entropy, namely $(1-entropy)$, leveraging on $entropy$ being between 0 and 1.

\begin{figure}[t]
    \centering
    \includegraphics[width=0.8\linewidth]{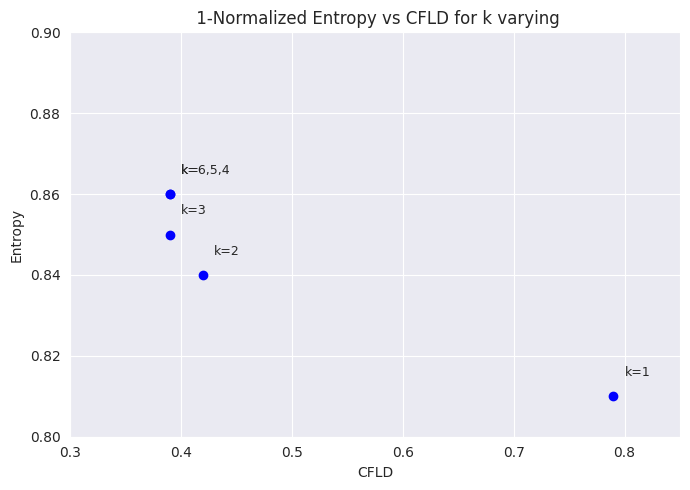}
    \caption{Average values between the generated log of the 8 case studies reported in Section~\ref{subsec:use_cases}, for different values of the parameter $k$. Different values of Trace Entropy are reported in the y-the axis, while different values of CFLD are reported in the x-axis.}
    \label{fig:k_cfld_fig}
\end{figure}

A common approach in multi-objective optimization is to identify the point that achieves the best trade-off between two different objectives, often visualized as the point closest to the origin in a two-dimensional metric space. This strategy, commonly referred to as the \textit{Elbow Method}, allows to select a hyperparameter value that balances competing criteria without favoring one over the other. Applying this procedure to our case, where the axes correspond to \textit{CFLD} and the complementary of normalized trace entropy 
$(1-entropy)$, the optimal hyperparameter value $k$ is determined as the one minimizing the Euclidean distance to the origin. The results are illustrated in a two-dimensional plot depicted in Figure~\ref{fig:k_cfld_fig}, while the numerical values are reported in Table~\ref{tab:k_cfld_tab}. Following this procedure, the optimal value for $k$ is 3, representing the best compromise between similarity and log variability.

\begin{table}[t]
    \centering
    \begin{tabular}{c|c}
        \textbf{k} & \textbf{Distance from (0,0)}\\
        \hline
        6 & 0.944\\
        5 & 0.944\\
        4 & 0.944\\
        3 & 0.935\\
        2 & 0.939\\
        1 & 1.131\\
    \end{tabular}
    \caption{Distance from (0,0) of the values reported in Figure~\ref{fig:k_cfld_fig}.}
    \label{tab:k_cfld_tab}
\end{table}


\end{document}